\documentstyle[preprint,eqsecnum,aps]{revtex}

\begin{document}
\input epsf

\tightenlines


\preprint{\vbox{\hbox{JLAB-THY-98-42}
}}

\title{Phenomenology of Large $N_c$ QCD\footnote{Lectures presented at
``11th Indian-Summer School of Nuclear Physics,'' sponsored by the
Institute of Nuclear Physics, \v{R}e\v{z} and the Faculty of
Mathematics and Physics of Charles University, Prague, Czech Republic,
September 7--11, 1998.}}

\author{Richard F. Lebed}

\vskip 0.1in

\address{Jefferson Lab, 12000 Jefferson Avenue, Newport News, VA
23606, USA\\E-mail: lebed@jlab.org}

\vskip 0.1in

\date{October, 1998}
\vskip .1in

\maketitle

\begin{abstract}
These lectures are designed to introduce the methods and results of
large $N_c$ QCD in a presentation intended for nuclear and particle
physicists alike. Beginning with definitions and motivations of the
approach, we demonstrate that all quark and gluon Feynman diagrams are
organized into classes based on powers of $1/N_c$.  We then show that
this result can be translated into definite statements about mesons
and baryons containing arbitrary numbers of constituents.  In the
mesons, numerous well-known phenomenological properties follow as
immediate consequences of simply counting powers of $N_c$, while for
the baryons, quantitative large $N_c$ analyses of masses and other
properties are seen to agree with experiment, even when ``large''
$N_c$ is set equal to its observed value of 3.  Large $N_c$ reasoning
is also used to explain some simple features of nuclear interactions.
\end{abstract}

\thispagestyle{empty}

\newpage
\setcounter{page}{1}

\section{Motivations and Fundamentals}
	The purpose of these lectures is to define, explain, and
exhibit the methods of large $N_c$ QCD for phenomenologists of both
the nuclear and particle bent.  Several excellent reviews on this
topic exist in the literature, each with a different emphasis, and to
each I am indebted for aspects of the pedagogy presented in the
current work.  Witten's article on large $N_c$ baryons\cite{Witt}, one
of my favorite papers, actually begins with a review of large $N_c$
mesons.  Chapter 8 in Coleman's {\it Aspects of Symmetry}\cite{Cole}
covers the development of large $N_c$ through 1980.  Many of the newer
developments with baryons are included in Manohar's 1997 Les Houches
lectures\cite{Man1} and Jenkins' review\cite{Jenk1}.

	I assume that initially many readers may find the notion of
large $N_c$ to be a rather exotic and unlikely approach to studying
the strong interactions, as it seemed to me when I began to learn it
some years ago.  As always with an unfamiliar topic, it is important
to start with clear definitions and motivations.  We begin with the
assertion that QCD-like theories possess a peculiar limit in which the
physics of the strong interactions becomes much simpler.  It is the
so-called {\it large $N_c$ limit}, in which the number of color
charges, $N_c$ (= 3 in our universe) is taken as a free parameter
({\it i.e.}, the gauge group is SU($N_c$)), and one considers the
limit $N_c \to \infty$.  Physical quantities are then considered in
this limit, where corrections appear at relative orders $1/N_c$,
$1/N_c^2$, {\it etc.}, the {\it $1/N_c$ expansion}.  Shortly after its
discovery by 't~Hooft\cite{tH1} in 1974, the method was applied to
mesons with many qualitative successes in explaining phenomenology.
However, it has only been in the past several years that much success
has come in the analysis of baryons, with much of it quantitative.

	However, before a detailed discussion of these very
interesting points, it is important to consider some simple but
incisive questions.  How does large $N_c$ work?  What does it predict?
And, perhaps most importantly, is it believable?

	In this section, we discuss how and why large $N_c$ should
work, and how it is implemented at the very basic level of QCD, in
terms of quarks and gluons.  To start with, the very notion of the
approach creates an interesting paradox: How can increasing the number
$N_c$ of degrees of freedom actually simplify the strong interaction
problem?  It is certainly true that the detailed interactions of any
particular quark or gluon become much more complicated if one insists
on solving the theory exactly.  To illustrate this point, let us
introduce two analogies; we shall come to appreciate that they
represent the way large $N_c$ works for mesons and baryons,
respectively.  The first analogy is that of a particle moving in a
complicated potential, one that requires many parameters for a
complete description, and the second is the many-body problem in
mechanics.  Both are difficult because of the large number of
available dynamical degrees of freedom.  However, if increasing the
number of degrees of freedom permits one to consider only {\em
collective\/} couplings in a systematic way, then the form of the
dynamics simplifies.  In the first analogy, the potential might be
decomposable into a simple dominant piece (which nonetheless
originates from a large number of more fundamental interactions),
while the other portions are subleading perturbations.  In the second,
statistical mechanics describes the system's gross features in terms
of just a few collective quantities, such as temperature or volume;
fluctuations about these values are highly suppressed.

	With these qualitative arguments in mind, what is special
about the large $N_c$ limit of QCD in particular?

\begin{enumerate}

\item {\em It is a perturbative QCD expansion.} Compare QCD to QED;
why does the latter work so much better?  Obviously, $\alpha_{\rm QED}
\sim 1/137$ is a perturbative expansion parameter, while $\alpha_{\rm
QCD} = O (1)$ at hadronic energy scales is clearly not perturbative.
At high energies, $\alpha_{\rm QCD}$ runs to smaller values due to
asymptotic freedom, but one would prefer a perturbative parameter that
works at {\em all\/} scales.  In fact, only one such parameter is
known to exist for QCD-like theories, namely, $1/N_c$.

\item {\em Physics simplifies in the large $N_c$ limit.}  As we show
presently, certain classes of Feynman diagrams dominate over others in
physical amplitudes, and this organization leads to many interesting
simplifications.  For example, we shall see that a color singlet $\bar
q q$ pair always appears at leading order in $1/N_c$ as a single
meson.

\item {\em It provides an explanation of phenomena hard or impossible
to explain in field theory.} For example, OZI suppression and the
physical dominance of resonant-mediated processes are consequences of
large $N_c$, as we shall see.

\item {\em It seems to work, even for $N_c = 3$.} This is perhaps the
most startling reason; indeed, one might expect that $1/N_c = 1/3$ is
too large to be a reasonable expansion parameter.  However, such an
objection may be answered in two ways.  One is described by
Coleman\cite{Cole} as Witten's ``wisecrack'': Consider $\alpha_{\rm
QED} = e^2 / 4\pi \sim 1/137$, which leads to $e \sim 0.3 \sim 1/3$.
If $1/N_c$ is too large an expansion parameter, why don't we say the
same thing about the QED charge $e$?  Turning the argument around, one
can find cases where the expansion parameter turns out to be $1/N_c^2
\sim 1/10$, which is more satisfactory to some critics of the
approach.  However, I prefer a much more pragmatic answer, which is
just that one should perform a given calculation with $N_c$ a free
parameter, set it to 3 at the end of the day, and see for oneself
whether or not the factors of 3 are supported by the experimental
measurements.  The ultimate justification of any approach in this view
is whether it successfully explains physical observations and can be
tested in new situations.

\end{enumerate}

	Before proceeding, consider the evidence that $N_c=3$ in our
universe; it is important to check that there is nothing fundamentally
immutable about this particular value that would impede consideration
of larger $N_c$.  Historically, the original reason for the invention
of color\cite{color} was to explain how fermionic baryons could have a
completely symmetric spin-flavor-space wavefunction,\footnote{Of
course, everything said here about flavor is equally true in the
two-flavor case of isospin.} as suggested by experiment.  A canonical
example is $\Delta^{++} (J_z = +3/2) = (u \!  \uparrow) (u \!
\uparrow) (u \!  \uparrow)$, which is completely symmetric under
spin-flavor indices.  Since the spatial wavefunction for this
low-lying state is expected to be completely symmetric, so is the
combined spin-flavor-space wavefunction, and therefore one requires a
new degree of freedom with (at least) 3 values completely
antisymmetric under quark exchange.

	The experimental confirmation of the color idea came with the
measurement of the ratio
\begin{equation}
R_{\rm hadron} \equiv \sigma(e^+ e^- \to {\rm hadrons}) / \sigma(e^+ e^-
\to \mu^+ \mu^-)
\end{equation}
away from thresholds for new flavor production, which is found, as
originally predicted\cite{CPT}, to be a factor of 3 larger than what
one would predict from a colorless quark theory.  Similarly,
perturbative QCD, in those high-energy kinematic regimes where it is
believed valid, only gives results in agreement with experiment when
one sets $N_c = 3$.  Furthermore, the calculated rate for $\pi^0 \to
\gamma \gamma$, which matches experiment, includes a factor of
$N_c=3$.

	Finally, the Standard Model is only renormalizable if the
particle content is chosen to eliminate the possibility of chiral
anomalies.  In particular, since quark charges are known from deep
inelastic scattering experiments to be integral multiples of 1/3, one
can show that chiral anomalies cancel only for the choice $N_c = 3$.
For universes with larger values of $N_c$, the cancellation of
anomalies is accomplished by adjusting the quark charges accordingly,
to be integral multiples of $1/N_c$.

	With these lessons in mind, let us consider the valence
structure of hadrons in large $N_c$.  In order to make progress, we
first make the crucial assumption that color confinement occurs for
arbitrarily large $N_c$; otherwise, such a universe bears no
resemblance to ours, since then free colored hadrons or even free
quarks can be abundant.  Such an assumption is not required, of
course, since the physical extrapolation from $N_c = 3$ to $N_c \to
\infty$ might lead from a confining to a nonconfining phase at some
value of $N_c >3$, but then the large $N_c$ limit would be of little
phenomenological value since we have no empirical knowledge of
nonconfining strong interactions.

	Mesons of valence flavor structure $\bar q_1 q_2$ with color
indices $\alpha$, $\beta$ for arbitrary $N_c$ have the normalized
wavefunction
\begin{equation} \frac{1}{\sqrt{N_c}} \,
(\bar q_1)_\alpha (q_2)^\beta \, \delta^\alpha{}_{\! \beta} ,
\end{equation}
where $\delta^\alpha{}_{\! \beta}$ serves to combine the $\bar q q$
pair into a color singlet.  The salient point is that mesons still
have the quantum numbers of a system with a single quark and a single
antiquark for arbitrary $N_c$.  Baryons, on the other hand, are built
in analogy with the $N_c=3$ observation that fermionic states with
completely symmetric spin-flavor-space wavefunctions require at least
as many colors as quarks.  On the other hand, once we specify that
there are $N_c$ colors, SU($N_c$) group theory requires a minimum of
$N_c$ colored quarks to form a color singlet.  Therefore, large $N_c$
baryons require exactly $N_c$ valence quarks.  Given the valence
quarks $q_1, q_2, \ldots, q_{N_c}$, the normalized wavefunction
including color indices $\alpha_1, \alpha_2, \ldots, \alpha_{N_c}$ is
\begin{equation}
\frac{1}{\sqrt{N_c !}} \, (q_1)^{\alpha_1} (q_2)^{\alpha_2} \cdots
(q_{N_c})^{\alpha_{N_c}} \, \epsilon_{\alpha_1 \alpha_2 \cdots
\alpha_{N_c}} .
\end{equation}
Furthermore, only odd $N_c$ produces fermionic baryons, since each of
the $N_c$ quarks has spin 1/2.  The large $N_c$ limit may be used
equally well to consider even-$N_c$ bosonic baryons, but of course
these are not phenomenologically relevant.

	We are now ready to consider large $N_c$ QCD itself, as
developed by 't~Hooft.  To the non-initiate who has been told of large
$N_c$ results but never taught its origins, the whole idea might still
seem a bit like necromancy.  But like any good magic show, the
essential predictive power of large $N_c$ just comes from trap doors
(topology) and mirrors (group theory).

\begin{figure}
  \begin{centering}
  \def\epsfsize#1#2{1.0#2}
  \hfil \epsfbox{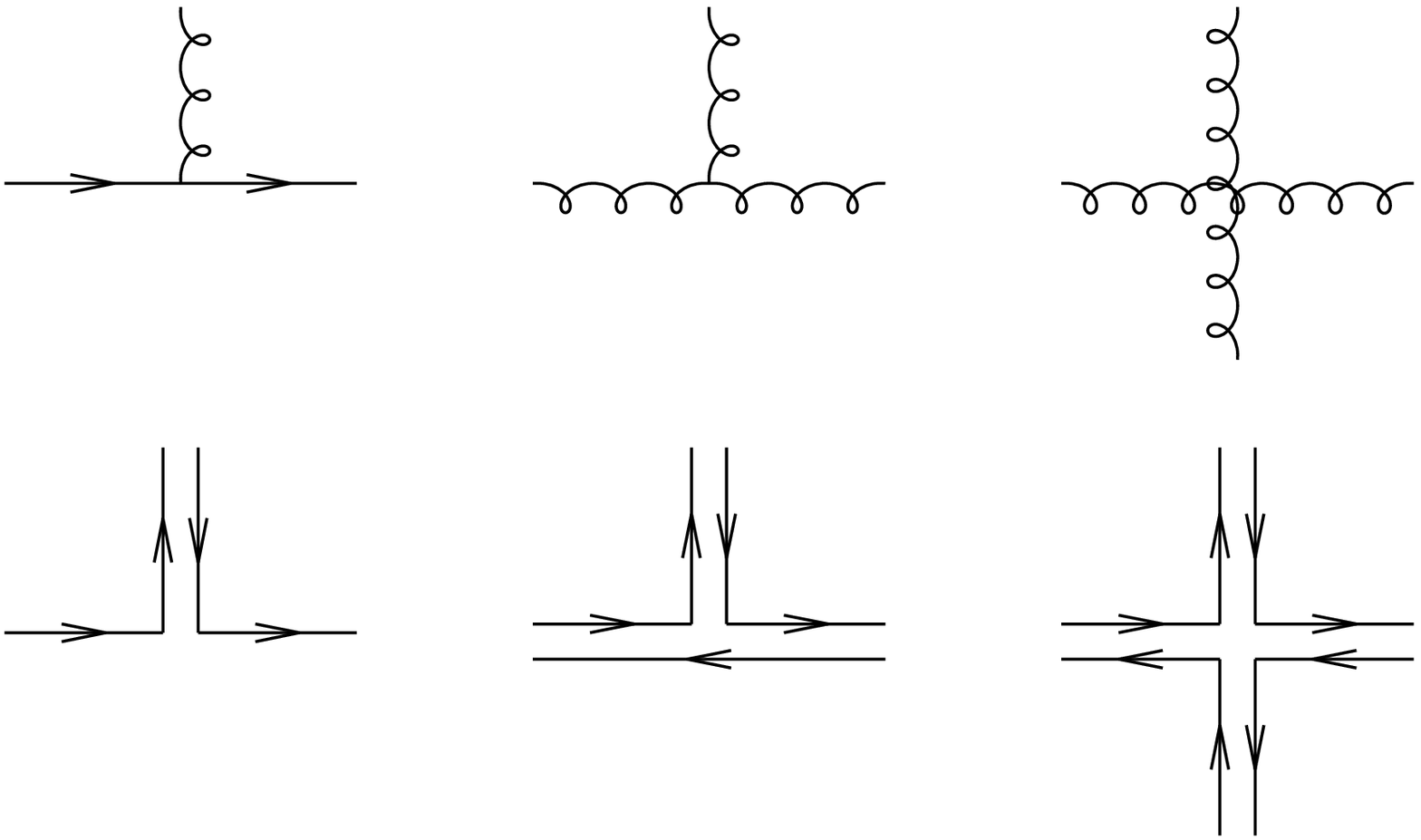}\hfil\hfil

\caption{The three types of QCD vertex in standard Feynman notation in
the top row, accompanied by their representations in double-line
notation in the bottom row.} \label{vert}

  \end{centering}
\end{figure}

	Let us begin with the usual particle content of QCD with the
color gauge group SU($N_c$).  Quarks carry a color index in the
fundamental representation of the group, which means that the standard
representation is a column vector $q^\alpha$ with $N_c$ components,
one for each color.  Diagrammatically, it is represented by an arrowed
line.  Antiquarks carry a color index in the fundamental conjugate, or
anti-fundamental, representation, and therefore appear in standard
form as a row vector $q_\alpha$ with $N_c$ components.  They are also
drawn as arrowed lines, but the (anti-) color flow is understood to be
opposite the direction of the arrow.  Gluons, which appear in the
adjoint representation of SU($N_c$), carry one color and one
anti-color index, and appear in standard form as traceless $N_c \times
N_c$ matrices $A^\alpha{}_{\! \beta}$, which can be drawn as two
parallel lines whose arrows point in opposite directions.  This is
called 't~Hooft's {\it double-line notation};\footnote{A couple of
notes for the inquisitive.  First, double-line notation works as well
for Faddeev-Popov ghosts (which also live in the adjoint
representation), so we may simply think of them in large $N_c$
counting as another type of gluon.  Second, the double-line notation
is strictly accurate for the gauge group U($N_c$) rather than
SU($N_c$).  What about the extra U(1) gluon?  Since it is only one
gluon out of $N_c^2$, one can show\cite{Cole,Man1} that it always
gives diagrams subleading by at least two powers of $N_c$.} it is
actually extraordinarily convenient in capturing the color physics,
because exact color conservation implies that color lines are
continuous, while the confinement assumption means that color lines
form only closed loops.  The three QCD vertices in double-line
notation are depicted in Fig.~\ref{vert}.

	The double-line notation is particularly useful as a guide to
developing an understanding of how the QCD gauge parameter, $g_s$ ($s$
= strong), scales as $N_c \to \infty$.  In fact, as 't~Hooft argued in
his original paper\cite{tH1}, QCD Green functions only have smooth,
finite limits as $N_c \to \infty$ if $g_s \propto 1/\sqrt{N_c}$.
Since this is such a central result in the analysis to come, we prove
the claim in detail below.  However, let us ponder for a moment why
such a result is interesting.  Given that $g_s \propto 1/\sqrt{N_c}$,
one can show that certain classes of Feynman diagrams dominate over
others, in possessing fewer powers of $1/N_c$, an organization known
as {\it large $N_c$ power counting}.  The dominant class of graphs is
denoted {\it planar}, which means that these graphs can be drawn in
the double-line notation in the plane of paper, with any quark line
(if one is present) placed on the perimeter, and with color lines
arranged to cross only at vertices.  Each nonplanarity, as will be
shown, costs a suppression factor of $1/N_c^2$.  Moreover, diagrams
with no internal quark loops dominate over the others, in that each
internal quark loop costs a factor of $1/N_c$.  So, in summary, large
$N_c$ counting provides an organization of all Feynman diagrams into a
well-defined and countable hierarchy of classes.

	We now present a proof of 't~Hooft's scaling of $g_s$.
Vertices typically fall into three types.  There are bilinears that
create $\bar q q$ pairs from the vacuum, and are accompanied by
$g_s^0$; let there be $V_2$ of these {\it current
insertions}.\footnote{In addition, the bilinear may also create $n$
gluons, for which gauge invariance requires including a factor
$g_s^n$, but such vertices can also be shown to satisfy the counting
rules.}  Each of $V_3$ trilinear ($\bar q qg$ or $ggg$) vertices
includes $g_s^1$, while each of $V_4$ quartic vertices includes
$g_s^2$.  Confinement requires that the $C$ color lines in the digram
completely close; then one may choose any of the $N_c$ colors to
appear on each loop, providing a combinatoric factor of $N_c^C$.  In
total, then, the diagram scales as
\begin{equation}
{\cal A} \propto g_s^{V_3+2V_4} N_c^C .
\end{equation}
Since the diagram has no external lines by color confinement, both
ends of each propagator $P$ terminate on a vertex:
\begin{equation}
2P = \sum_n nV_n = 2V_2 + 3V_3 + 4V_4 .
\end{equation}
Defining the total number of vertices
\begin{equation}
V = \sum_n V_n = V_2 + V_3 + V_4,
\end{equation}
one has
\begin{equation} \label{ncount}
{\cal A} \propto g_s^{2P-2V} N_c^C .
\end{equation}
It is very fruitful to use the double-line notation to interpret the
diagram as a polyhedron, much like a geodesic roof.  Each color loop
$C$ becomes a {\it face}, where each propagator $P$ is an {\it edge\/}
and each vertex $V$ of the diagram is also a {\it vertex\/} of the
solid figure.  Then each quark loop, which possesses one less color
line than if it were replaced by a (double-line) gluon loop, may be
thought of as a missing face, or topological {\it boundary\/} $B$,
while each nonplanar line may be interpreted as a handle (since it
represents one edge forced to pass in front of another but not meet at
a vertex), or topological unit of {\it genus\/} $G$.  We relate these
features to topological objects so that we may now use a famous
theorem due to Euler: For each
orientable\footnote{\label{orient}Double-line diagrams are orientable,
{\it i.e.}, possess a well-defined and continuous definition of normal
vector to their surfaces, and hence, an inside and outside.  This is
because each double-line edge has one arrow pointing in either
direction: Recall that such a feature is used when proving Stokes'
theorem in calculus.  In turn, this feature of the double-line
notation comes from using SU($N_c$) as a gauge group, where gluons may
be thought of as having (quark + antiquark) quantum numbers.  The same
would not be true in, {\it e.g.}, O($N_c$).}  figure, there is a
topological invariant called the Euler character $\chi_E$ satisfying
\begin{equation} \label{euler}
\chi_E = C - P + V = 2 - 2G - B .
\end{equation}
Therefore,
\begin{equation}
{\cal A} \propto (g_s^2 N_c)^C \left[ (g_s^2 N_c)^{-1} \, N_c
\right]^{2-2G-B} .
\end{equation}
The first term still depends on the particular diagram, while the
second has a purely topological power.  Now observe that, if $g_s$
falls off faster than $1/\sqrt{N_c}$, then diagrams of any given
topology with the fewest color loops, and hence a minimal number of
gluons, dominate; but this is a trivial theory, in that only valence
diagrams are important.  Clearly the gluon degrees of freedom are
vital to understanding the strong interaction, so we reject this
possibility.  On the other hand, if $g_s$ falls off slower than
$1/\sqrt{N_c}$, then diagrams of any given topology with the most
color loops dominate; but this is a nonsense theory, for one can
always add more color loops to diagrams of a given topology, and in
this scenario one could make no predictions at all.  However, we know
that this is not the case, since perturbative QCD in high-energy
regimes works well.  We conclude that $g_s \propto 1/\sqrt{N_c}$ is
the only nontrivial, sensible large $N_c$ scaling,\footnote{Yet again,
nature might not require ``real'' large $N_c$ QCD to satisfy the
conditions of nontriviality and sensibility, but if it does not, one
cannot extrapolate from $N_c \to \infty$ back to $N_c = 3$ and hope to
make predictions.} giving the result
\begin{equation}
{\cal A} \propto N_c^{2-2G-B},
\end{equation}
meaning that the large $N_c$ counting of any particular diagram is
purely topological.  Since this is the case, for any given diagram one
may add an arbitrarily large number of planar gluons and remain within
the same topological class, {\it i.e.}, the diagram maintains the same
large $N_c$ counting.  However, each additional quark loop or
nonplanar gluon adds one unit to $B$ or $G$, leading to an additional
$1/N_c$ or $1/N_c^2$ suppression, respectively.

	Another (although actually equivalent) way to demonstrate $g_s
\propto 1/\sqrt{N_c}$ is to start with the QCD renormalization group
equation,
\begin{equation} \label{RGE}
\mu \frac{dg_s(\mu)}{d\mu} = - b_0 \frac{g_s (\mu)^3}{16\pi^2} +
O(g_s^5(\mu)) ,
\end{equation}
where $b_0 = (11N_c - 2N_f)/3 > 0$, with $N_f$ the number of light
quark flavors, and the positivity of $b_0$ being the reason we know
that QCD is asymptotically free.  Let us rescale $g_s = \bar g_s
N_c^{-1/2 + \epsilon}$, with $\bar g_s$ finite as $N_c \to \infty$.
Then (\ref{RGE}) reads
\begin{equation}
\mu \frac{d\bar g_s (\mu)}{d\mu} = -\frac{1}{3} \left( 11 -
2\frac{N_f}{N_c} \right) \frac{\bar g_s^3 (\mu)}{16 \pi^2}
N_c^{2\epsilon} + O(\bar g_s^5 (\mu)) .
\end{equation}
We immediately see that, if $\epsilon <0$, $\bar g_s$ is constant and
$g_s$ instantly runs to zero for $N_c$ large (assuming that the theory
has a perturbative limit), leading to a trivial theory of valence
diagrams only; on the other hand, if $\epsilon >0$, $\bar g_s$ runs
infinitely fast, making a nonsense theory with no meaningful
predictions.  Again, we conclude that only $g_s \propto 1/\sqrt{N_c}$
leads to a nontrivial and sensible theory.

	We conclude this section by illustrating in Figs.~\ref{topo1}
and \ref{topo2} the large $N_c$ counting in the three different
approaches: counting powers of $g_s$ and $N_c$, counting features of
the equivalent polyhedra, and counting topological invariants.  These
examples should provide the reader with hands-on experience with the
theorems we have considered.

\section{Mesons in the $1/N_c$ Expansion} \label{mes}

	Having shown that the large $N_c$ limit of QCD is only
nontrivial and sensible if $g_s \propto 1/\sqrt{N_c}$, and that
Feynman diagrams involving quarks and gluons fall into classes labeled
by powers of $N_c$ with planar diagrams at leading order, what have we
gained?  Certainly, the observed strongly interacting particles are
hadrons rather than quarks and gluons, and so if there were no obvious
relation between the two pictures --- beyond the valence quark model,
which unlike large $N_c$ contains no explicit gluons --- large $N_c$
would hold little phenomenological interest.  But in fact one can
derive an elegant and simple large $N_c$ relationship between hadrons
and their fundamental QCD constituents.  In this section, we consider
this derivation and its consequences for mesons.

	A good starting point for these discussions is to consider the
particle spectrum of large $N_c$.  In our $N_c=3$ world, one observes
{\it conventional mesons\/} with the quantum numbers of $\bar q q$,
baryons with the quantum numbers of $qqq$, and more recently, {\it
hybrid mesons}\cite{E852} with the quantum numbers of ($\bar q q$ +
gluons).  On the other hand, one might equally well expect to see {\it
glueballs}, which are defined as mesons containing all gluons and no
valence quarks, exotic {\it multiquark mesons\/} with the quantum
numbers of ($\bar q q$ + extra $\bar q q$ pairs), {\it exotic
baryons\/} with the quantum numbers of ($qqq$ + extra $\bar q q$
pairs), and {\it hybrid baryons}, with the quantum numbers of ($qqq$ +
glue).  Members of the latter class have not yet been observed, and
with the exception of hybrid baryons, large $N_c$ provides an
explanation.

\begin{figure}
  \begin{centering}
  \def\epsfsize#1#2{0.6#2}
  \hfil \epsfbox{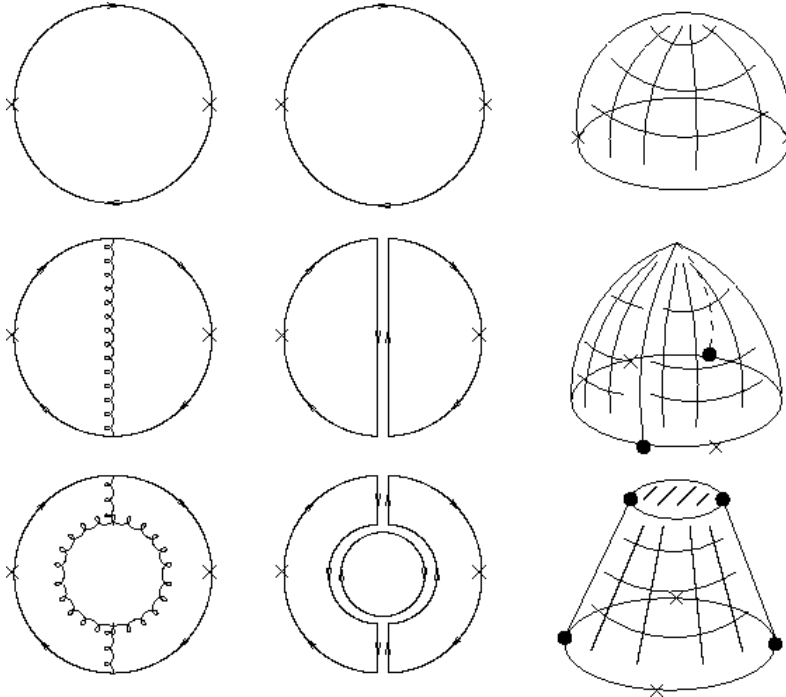} \hfil\hfil

\caption{Examples of large $N_c$ counting for Feynman diagrams.  Each
row gives three equivalent representations of the same diagram, the
first in standard form, the second in double-line notation, and the
third as a topological figure.  In terms of the quantities in
Eqs.~(\ref{ncount}) and (\ref{euler}), the first diagram has $g_s^0
N_c^1$, $C=1$, $P=2$, $V=2$, $G=0$, $B=1$; the second has $g_s^2
N_c^2$, $C=2$, $P=5$, $V=4$, $G=0$, $B=1$; the third has $g_s^4
N_c^3$, $C=3$, $P=8$, $V=6$, $G=0$, $B=1$.  All three scale as $N_c^1$
and are planar.  Crosses indicate current insertions connecting quarks
to the QCD vacuum.} \label{topo1}

  \end{centering}
\end{figure}

\begin{figure}
  \begin{centering}
  \def\epsfsize#1#2{0.5#2}
  \hfil \epsfbox{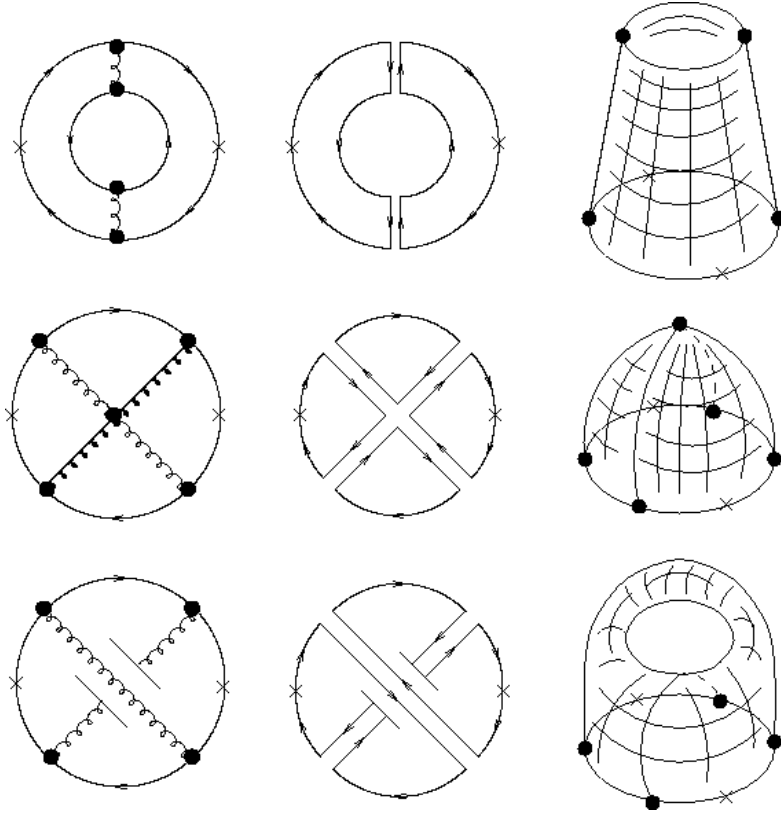} \hfil\hfil

\caption{More examples of large $N_c$ counting for Feynman diagrams.
The first line has $g_s^4 N_c^2$, $C=2$, $P=8$, $V=6$, $G=0$, $B=2$
and scales as $N_c^0$; the second has $g_s^6 N_c^4$, $C=4$, $P=10$,
$V=7$, $G=0$, $B=1$ and scales as $N_c^1$; the third has $g_s^4
N_c^1$, $C=1$, $P=8$, $V=6$, $G=1$, $B=1$, and scales as $N_c^{-1}$.
Note that only the second diagram is planar and hence leading in
$1/N_c$.} \label{topo2}

  \end{centering}
\end{figure}

	To see how it works, first consider QCD two-point diagrams as
depicted in Figs.~\ref{topo1}, \ref{topo2}.  The current $J$ creates a
$\bar q q$ pair from the QCD vacuum at some spacetime point, the
quarks propagate and interact for some interval, and then the pair is
annihilated by a current $J^\dagger$.  We ask, what kinds of hadrons
can appear in the intermediate state?  Equivalently, using our
confinement assumption, we want to discover what color-singlet
combinations of quarks and gluons are revealed if we cut the diagram
at any convenient place.  We claim that, to leading order in $1/N_c$,
cutting such a diagram leads only to a single $\bar q q$ meson --- no
multiquark mesons, no glueballs.

	The proof of this statement relies on planar diagrams with no
internal quark loops (which alone eliminates multiquark mesons from
our discussion) dominating in the $1/N_c$ expansion.  Typical planar
diagrams in the double-line notation are those exhibited in
Fig.~\ref{topo1}.  The unique feature possessed by planar diagrams and
no others is that, when cut, each color index appears adjacent to its
corresponding anticolor index.  That is, one cuts through an entire
color loop before passing to the next one.  In terms of fields, the
equivalent (perhaps nonlocal) operator has color structure
\begin{equation}
\bar q_{\alpha_1} A^{\alpha_1}{}_{\! \alpha_2} A^{\alpha_2}{}_{\!
\alpha_3} \cdots A^{\alpha_{n-1}}{}_{\! \alpha_n} \, q^{\alpha_n}.
\end{equation}
In terms of the generators of color indices, one may use the usual
Gell-Mann matrices $\lambda^a$ to expose the underlying structure
\begin{equation}
\left( \lambda^{a_1} \right)^{\alpha_1}{}_{\! \! \! \alpha_2} \left(
\lambda^{a_2} \right)^{\alpha_2}{}_{\! \! \! \alpha_3} \cdots \left(
\lambda^{a_n} \right)^{\alpha_{n-1}}{}_{\! \! \! \alpha_n} .
\end{equation}
Now one uses a little SU($N_c$) group theory to complete the proof.  A
particularly convenient way to do this is to exhibit the commutation
relations obeyed by the generators:
\begin{equation}
\left[ \lambda^a , \lambda^b \right] = 2if^{abc} \lambda^c ,
\hspace{2em} \left\{ \lambda^a , \lambda^b \right\} = \frac{4}{N_c}
\delta^{ab} + 2d^{abc} \lambda^c .
\end{equation}
Combining these,
\begin{equation}
( \lambda^a )^{\alpha_1}{}_{\! \alpha_2} ( \lambda^b
)^{\alpha_2}{}_{\! \alpha_3} = \frac{2}{N_c} \delta^{ab}
\delta^{\alpha_1}{}_{\! \alpha_3} + \left( d^{abc} + if^{abc}
\right) ( \lambda^c )^{\alpha_1}{}_{\!  \alpha_3} ,
\end{equation}
where the first term, which is color singlet, is suppressed by
$1/N_c$, while the second term, which is color adjoint like the
original gluons, is of leading order, $O(N_c^0)$.  By induction, every
string of gluon fields is color adjoint to leading order, and
therefore the only way to make a leading-order color singlet from the
diagram is to take the operator $\bar q AA \cdots Aq$ as an
irreducible whole.  The conclusion is that only a single meson, and no
glueballs or multiquark mesons, appear at leading order in $1/N_c$.

	This important result can now be used to prove a wide variety
of consequences.  Since each planar two-point diagram with a $\bar q
q$ pair appears as a single meson, the sum of planar two-point
diagrams with momentum transfer $p^2$ must also equal a sum of meson
propagators,
\begin{equation}
\mbox{\rm Planar diagram sum} = O(N_c^1) = \sum_n
\frac{f_n^2}{p^2-m_n^2} ,
\end{equation}
where $f_n$, the $n$th meson decay constant, is defined as the
transition amplitude between meson state $n$ and the vacuum through
the current $J$.  Using that $p^2$ is arbitrary and that the equality
must hold for any (sufficiently large) value of $N_c$, one finds that
$m_n^2 = O(N_c^0)$ and $f_n = O(\sqrt{N_c})$.  Furthermore, for $p^2$
very large, the diagram must approach its perturbative QCD result,
$\sim \ln (p^2/\mu^2)$, where $\mu$ is the renormalization point.
However, no finite sum of factors $f_n^2/(p^2-m_n^2)$ produces a
logarithm in $p^2$, so the number of meson states in large $N_c$ must
be infinite.  Since one may restrict this calculation to diagrams of a
given fixed $J^{PC}$, it is also true that the number of mesons of any
particular quantum numbers is also infinite.

\begin{figure}
  \begin{centering}
  \def\epsfsize#1#2{2.0#2}
  \hfil \epsfbox{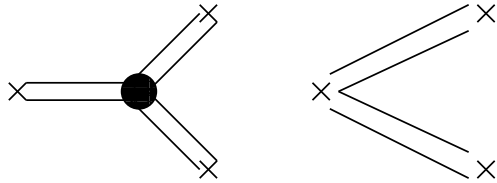} \hfil\hfil

\caption{The two distinct types of meson structures equivalent to the
QCD three-point function.} \label{3pt}

  \end{centering}
\end{figure}

	One may continue in this fashion to consider three-, four-,
and higher-point functions.  The diagrams in these cases resemble
those of the two-point function, except for the inclusion of more
current insertions, and thus are all $O(N_c^1)$.  However, unlike in
the two-point diagram, there is now more than one way to collect the
quark and gluon fields into color singlets.  For example, for the
three-point function, each of the three current insertions may create
a meson, which annihilate at a trilinear meson vertex
(Fig.~\ref{3pt}).  Alternately, one current insertion may create two
mesons, which propagate to the remaining two insertions.  The
three-point diagram is then given by
\begin{equation}
\sum_{1,2,3} \frac{f_1 f_2 f_3 \cdot (\mbox{\rm 3-meson
vertex})}{(p_1^2-m_1^2)(p_2^2-m_2^2)(p_3^2-m_3^2)} + \sum_{2,3}
\frac{(\mbox{\rm Amplitude to create 2 mesons}) \cdot f_2
f_3}{(p_2^2-m_2^2)(p_3^2-m_3^2)} .
\end{equation}
Using that $m_n^2 \sim N_c^0$, $f_n \sim \sqrt{N_c}$, one
discovers that the 3-meson vertex scales as $1/\sqrt{N_c}$, while the
amplitude to create 2 mesons scales as $N_c^0$.

	Diagrams with more current insertions allow for more types of
internal meson vertices, and the $N_c$ scaling behavior of each type
of vertex can be derived inductively.  For example, meson vertices
from the four-point function can be obtained by using results from the
two- and three-point functions.  In general, given an $m$-point
function, each possible $n$-point diagram, $n < m$, is required to
appear with the large $N_c$ scaling derived in a previous induction
step (a statement of unitarity) with all possible permutations of
meson propagators (a statement of crossing symmetry).  One finds in
this way the general results that both the $m$-meson vertex and the
amplitude to create $m$ mesons from the vacuum scale as $N_c^{1-m/2}$.
This important result appears to have been first obtained by
Veneziano\cite{Ven}.

	Such a scaling implies immediately that mesons are free, since
the propagator, {\it i.e.}, the ``two-meson vertex,'' scales as
$N_c^0$, and therefore neither blows up nor vanishes as $N_c \to
\infty$, while the $1/\sqrt{N_c}$ scaling of the three-meson vertex,
and the additional $\sqrt{N_c}$ suppression for each additional meson,
implies that mesons are stable --- a remarkable result!  But, could
mixing with possible exotic states spoil this reasoning?  We now
discuss the fate of exotics to show that this does not occur.

	First consider glueballs; for simplicity, depict those with
only two valence gluons, whose two-point diagrams resemble those of
Figs.~\ref{topo1} and \ref{topo2}, except that the exterior quark is
replaced by a gluon.  Then all the large $N_c$ counting is the same as
before, except that the single color line of the quark is replaced by
the double line of the gluon.  Therefore, each diagram is a factor of
$N_c$ larger, so that the planar diagrams scale as $N_c^2$.  The same
proof as for mesons shows that these diagrams are dominated by the
one-glueball state, and so one finds that the masses and decay
constants of glueballs scale as $N_c^0$ and $N_c^1$, respectively.
Continuing with the same logic as before, the $g$-glueball vertex
scales as $N_c^{2-g}$.

\begin{figure}
  \begin{centering}
  \def\epsfsize#1#2{0.7#2}
  \hfil \epsfbox{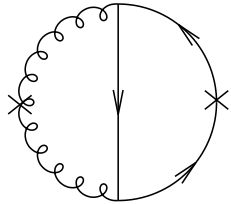} \hfil\hfil

\caption{One leading diagram for glueball-meson mixing.}
\label{gluemesmix}

  \end{centering}
\end{figure}

	As for glueball-meson mixing, the simplest planar diagram for
the process is exhibited in Fig.~\ref{gluemesmix}.  Converting it to
double-line format, we find two color loops and two factors of $g_s$,
so that the diagram scales as $N_c^1$.  Using our previous results for
the glueball and meson decay constants of $N_c^1$ and $\sqrt{N_c}$,
which appear at the current insertions, we see that the glueball-meson
coupling scales as $1/\sqrt{N_c}$, meaning that glueballs and mesons
do not mix at leading order.  Therefore, glueballs and mesons are {\em
separately\/} free, stable particles as $N_c \to \infty$, with the
combined result that the $g$-glueball, $m$-meson vertex for $m >0$
scales as $N_c^{1-g-m/2}$.

\begin{figure}
  \begin{centering}
  \def\epsfsize#1#2{0.7#2}
  \hfil \epsfbox{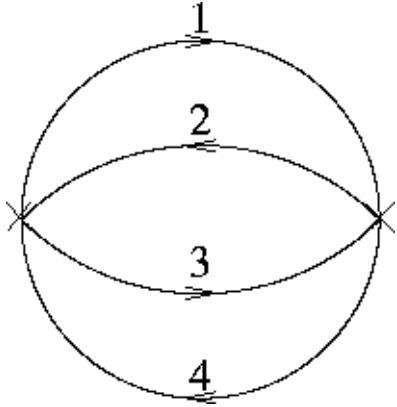} \hfil\hfil

\caption{Two-point diagram used to study production of multiquark
mesons.}
\label{multi}

  \end{centering}
\end{figure}

	Next consider multiquark mesons, {\it e.g.}, those with
valence quantum numbers $\bar q q \bar q q$.  These are created by
current insertions of the form $J = \bar q q \bar q q (x)$, as
depicted in Fig.~\ref{multi}.  To leading order in $1/N_c$, the color
loops close either as (1,2), (3,4) or (1,4), (2,3).  This means that
the diagram has two distinct closed color loops and scales as $N_c^2$;
however, we know from our previous reasoning that each such loop is a
single meson to leading order in $1/N_c$.  To create a single meson
with $\bar q q \bar q q$ quantum numbers requires an {\em
irreducible\/} color loop, such as (1,2,3,4) or (1,4,3,2); however,
such a diagram scales as $N_c^1$ and is thus subleading.  We conclude
that multiquark mesons are suppressed in $1/N_c$.

\begin{figure}
  \begin{centering}
  \def\epsfsize#1#2{1.0#2}
  \hfil \epsfbox{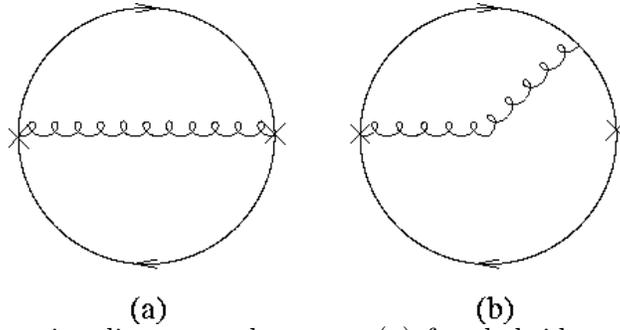} \hfil\hfil

\caption{Leading two-point diagrams relevant to $(a)$ free hybrid
meson propagation, and $(b)$ hybrid-conventional meson mixing.}
\label{hyb}

  \end{centering}
\end{figure}

	Finally, we turn our attention to hybrid mesons.  From the
previous paragraphs, one might expect that we shall dispose of these
as well.  But, in fact, they are present in the large $N_c$
limit\cite{Cohen}.  To see this diagrammatically, consider hybrids of
valence structure $\bar q q g$; one simple planar diagram representing
their propagation is given in Fig.~\ref{hyb}$a$.  One finds two color
loops and two factors of $g_s$, meaning that such diagrams scale as
$N_c^1$, the same as the leading conventional valence meson diagrams
as in Figs.~\ref{topo1}, \ref{topo2}.  In particular, one finds that
hybrid mesons have masses of $O(N_c^0)$ and decay constants of
$O(\sqrt{N_c})$, exactly like those of conventional mesons.  As for
meson-hybrid mixing, one leading diagram is depicted in
Fig.~\ref{hyb}$b$, which has two color loops and two factors of $g_s$,
thus scaling as $N_c^1$.  Of course, the symmetries of spacetime
require that the conventional and hybrid mesons in this case must
share the same quantum numbers.  Using that both conventional and
hybrid mesons have decay constants scaling as $\sqrt{N_c}$, one sees
that the meson-hybrid coupling is $O(N_c^0)$, so that the two mix
freely in large $N_c$.  An equivalent observation is that {\em all\/}
mesons in large $N_c$ are expected to have a finite hybrid component,
and the large $N_c$ limit sheds no light on the mystery of why almost
all observed mesons have only the quantum numbers of conventional
mesons.

	It should also be noted that hybrids with exotic (non-$\bar q
q$) quantum numbers (such as the newly-observed $1^{-+}$) do not mix
with conventional mesons at leading order in $1/N_c$, but are
nonetheless not hard to create.  The leading three-point diagrams
allow for one or more of the current insertions to have exotic quantum
numbers, which means that the process of exotic meson production
occurs at the same order in $N_c$ as that of conventional mesons.
Glueballs, on the other hand, are harder to create from meson sources
and slower to decay in the large $N_c$ limit; we leave the
verification of these facts as an opportunity for the reader to
experiment with double-line diagrams.

	In summary, our analysis of the large $N_c$ meson spectrum
teaches that, for the phenomenology of conventional mesons, one may
ignore glueballs, multiquark states, and hybrids with exotic quantum
numbers.  On the other hand, hybrids with conventional quantum numbers
should be treated just like conventional valence mesons.  The general
multiparticle vertex for $g$ glueballs, $m$ mesons, and $h$ hybrids
scales as
\begin{equation}
N_c^{1-g-m/2-h/2} ,
\end{equation}
for $m>0$ or $h>0$.  When $m=h=0$, the exponent is $2-g$.  The
physical mesons, which contain a hybrid component, are {\em free},
since the propagator scales as $N_c^0$; {\em stable}, since the
3-meson vertex is $O(1/\sqrt{N_c})$, and {\em non-interacting}, since
meson-meson scattering, {\it i.e.} the 4-meson vertex, is $O(1/N_c)$.

	In general, one concludes that diagrams with the fewest mesons
allowed tend to dominate physical processes.  This immediately gives a
field-theoretic explanation of the dominance of resonance-mediated
decays and the existence of excited mesons narrow enough to measure.
Although one could imagine an alternative strong interaction in which
every mesonic process with a substantial amount of energy release
simply produced a spray of pions and nothing else, the real strong
interaction very definitely produces a rich and complex spectrum of
observable resonances.  Large $N_c$ QCD gives an extraordinarily
economical explanation for this basic observation.

	One success of the large $N_c$ limit is the OZI
rule\cite{OZI}, that meson processes requiring the annihilation or
$\bar q q$ pairs, or equivalently, the presence of an all-glue
intermediate state, are phenomenologically suppressed.  The canonical
example is the decay of the $\phi (1020)$ meson, which is believed to
have the valence structure $\bar s s$.  It decays into a $K \bar K$
pair, for which there is almost no phase space ($2m_K \sim 1$ GeV)
83\% of the time, but only 17\% of the time into $\rho \pi$, $\pi \pi
\pi$, and other nonstrange modes, for which there is much more
available phase space.  Apparently, $\bar s s$ annihilation is
suppressed, and large $N_c$ offers a very simple explanation: A decay
like $\phi \to \bar K K$ may be represented with a standard,
leading-order ($O(N_c^1)$) three-point diagram as we have discussed,
but annihilating the strange quark closes its quark color loop and
requires another to create the light hadrons in the final state.
Since each quark loop costs a factor of $1/N_c$, the leading
OZI-suppressed diagrams inherit this suppression.\footnote{The case
with finite $N_c=3$ is much more subtle, as real OZI suppression
appears to have interesting dynamical origins above and beyond the
mere factor of $1/N_c = 1/3$: See \cite{GI}.}

	The observed fact that mesons appear in nonets of flavor SU(3)
rather than octets, {\it i.e.}, that the flavor octet and singlet
mesons tend to mix and have comparable masses, is another success of
large $N_c$.  The canonical example is $\eta (547)$ and $\eta^\prime
(958)$, which are the mass eigenstates of mixing between a flavor
singlet $\eta_1$ and flavor octet $\eta_8$; {\it a priori}, the
$\eta_1$ could have turned out much heavier than $\eta_8$.  However,
one can show that diagrams differentiating between their masses either
break SU(3), large $N_c$, or both.  To see this, begin in the SU(3)
flavor limit where $u$, $d$, and $s$ quarks are not distinguished.
Then all planar two-point diagrams such as in Fig.~\ref{topo1}
contribute equally to both masses, but a diagram with pure glue
intermediates contributes to $m^2 (\eta_1)$ alone, since gluons, like
$\eta_1$, are flavor singlets.  Because the latter diagrams are
suppressed by $1/N_c$, one concludes that terms symmetric in SU(3)
that distinguish the masses --- which {\it a priori\/} could have been
large --- are in fact suppressed by $1/N_c$ effects:
\begin{equation}
m^2 (\eta_1) - m^2 (\eta_8) = O \left( 1/N_c, \, m_s \right) .
\end{equation}

	As a final example of mesonic applications of large $N_c$,
consider the famous 't~Hooft model\cite{tHmodel}.  By definition, the
't~Hooft model is large $N_c$ QCD in 1 space and 1 time dimension
(1+1).  Although it is called a model, it is actually a full-fledged
quantum field theory that is exactly soluble, in that all hadronic
Green functions may be obtained entirely in terms of quark degrees of
freedom.  For example, one can compute meson masses precisely in terms
of quark masses.  Furthermore, the only asymptotically free states in
the theory are confined, color-singlet hadrons, which appear in an
infinite tower of increasing masses.

	So, although we assumed confinement in large $N_c$ QCD in 3+1
dimensions, it appears to be an immediate consequence in 1+1.  Such a
result might be expected based upon naive dimensional analysis.  For,
consider the momentum space potential derived from a single gluon
exchange propagator in the instantaneous approximation:
\begin{equation}
V(\mbox{\boldmath $q$}) \sim 1/\mbox{\boldmath $q$}^2 ,
\end{equation}
and Fourier transform to position space.  In 3+1 dimensions, one has
$V(\mbox{\boldmath $r$}) \sim 1/r$, the usual Coulomb interaction,
which is asymptotically free as $r \to \infty$; in 1+1,
$V(\mbox{\boldmath $r$}) \sim r$, which confines as $r \to \infty$.
That the confining potential survives relativistic and field theory
corrections is, however, nontrivial.

\begin{figure}
  \begin{centering}
  \def\epsfsize#1#2{2.0#2}
  \hfil \epsfbox{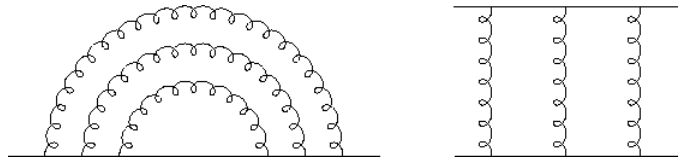} \hfil\hfil

\caption{Nontrivial ``rainbow'' and ``ladder'' diagrams summed by the
't~Hooft model.}
\label{tHdiag}

  \end{centering}
\end{figure}

	The 't~Hooft model is soluble precisely because of the two
features of its definition.  Large $N_c$ guarantees that nonplanar
gluons and virtual quark loops are suppressed, while working in 1+1
allows one to choose gauges for the gluon field such that gluon-gluon
couplings vanish.  Such terms arise in minimal substitution from the
commutator $[A_\mu, A_\nu]$ term; working in a gauge such that one of
the two components of $A$ vanishes guarantees that the commutator does
also.\footnote{Furthermore, if one chooses a linear gauge, then
Faddeev-Popov ghosts also vanish.}  The only remaining diagrams are
``rainbow'' and ``ladder'' types (Fig.~\ref{tHdiag}) and these can be
summed using ``bootstrap''-type Schwinger-Dyson equations, essentially
the same trick that is used to sum the geometric series.

	Performing this summation for the Green function of a $\bar q
q$ system, one obtains the meson wavefunction $\phi$ in terms of the
{\it 't~Hooft equation},
\begin{equation}
\mu_n^2 \phi_n^{M\overline{m}} (x) = \left( \frac{M_R^2}{x} +
\frac{m_R^2}{1-x} \right) \phi_n^{M\overline{m}}(x) - \int^1_0 dy \,
\phi_n^{M\overline{m}} (y) \, \Pr \frac{1}{(y-x)^2},
\end{equation}
where the available kinematic invariants are the quark masses $M$ and
$m$, the resultant meson mass eigenvalue $\mu$, and the 1+1 analogue
$x$ of the deep inelastic variable $x_{\rm BJ}$, defined by
\begin{equation}
x \equiv \frac{p_Q^0 + p_Q^1}{p_{\rm mes}^0 + p_{\rm mes}^1} \in [0,1]
.
\end{equation}
Qualitatively, $x$ and $1-x$ represent the fraction of meson momentum
carried by the quark and antiquark, respectively.  The quark masses
are renormalized by
\begin{equation}
m_R^2 = m^2 - \frac{g_s^2 N_c}{2\pi} ,
\end{equation}
from which we note that the 1+1 strong coupling $g_s$ actually has
dimensions of mass, and using that $g_s \sim 1/\sqrt{N_c}$, it is
natural to describe masses in units of $g_s \sqrt{N_c/2\pi}$.  Such a
quantity acts in the same way in 1+1 as $\Lambda_{\rm QCD}$ in 3+1, in
that it distinguishes ``heavy'' from ``light'' quarks.  The 't~Hooft
equation is known to possess an infinite tower of solutions, labeled
by $n = 0,1,2,\ldots$, which alternate in parity ($P = (-1)^{n+1}$).
As $n \to \infty$, the meson masses scale as $\mu_n^2 \to \pi^2 n$.
Since $x$ is bounded between 0 and 1, the 't~Hooft equation is
qualitatively similar to the quantum mechanical problem of a particle
in a box.

\begin{figure}
  \begin{centering}
  \def\epsfsize#1#2{0.5#2}
  \hfil \epsfbox{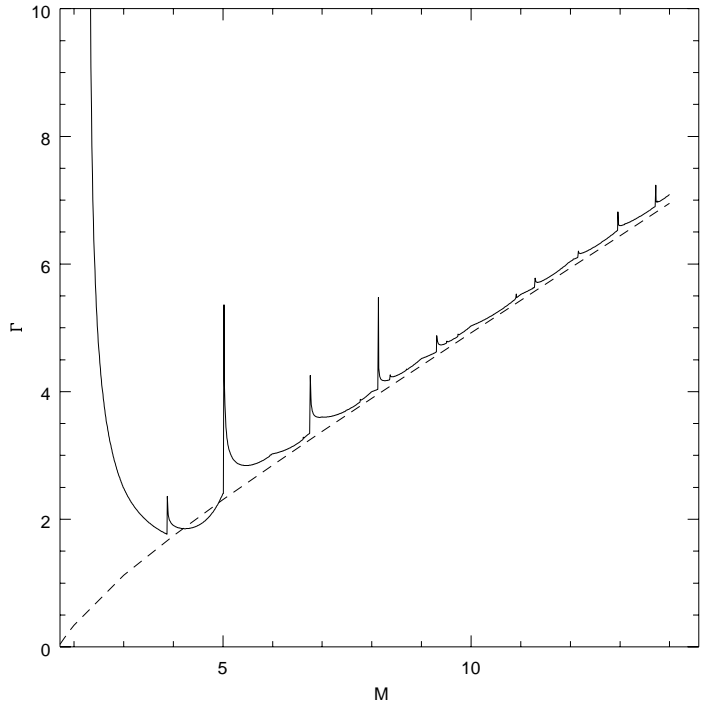} \hfil

\caption{Full rate for nonleptonic weak decay of meson containing a
heavy quark of mass $M$ as a function of $M$.  The solid line is
computed in the 't~Hooft model and sums over all allowed exclusive
modes, while the dashed line is due only to the free heavy quark weak
decay diagram.} \label{GLfig}

  \end{centering}
\end{figure}

	The utility of the 't~Hooft model is that it can be used to
study difficult problems of QCD that have not been fully solved in
3+1, such as the meson form factors, deep inelastic scattering, and
the saturation of hadronic rates by partonic diagrams.  For example,
consider\cite{GL} the nonleptonic weak decay of a meson containing a
heavy quark into lighter mesons.  As the heavy quark mass $M$ is
increased, the light antiquark and gluon degrees of freedom should
exert less and less influence on the decay of the heavy quark, and
eventually the process should be well described by the diagram of {\em
free\/} heavy quark weak decay.  Since one can compute both the
hadronic and partonic decay rates in the 't~Hooft model, the two can
be compared directly.  The result is Fig.~\ref{GLfig}, in which one
sees that the two rates rapidly approach one another for large $M$.

\section{Baryons in the $1/N_c$ Expansion}

	Recalling from Sec.~I that baryons for quarks of $N_c$ colors
have the quantum numbers of $N_c$ quarks, one is immediately faced
with the dilemma of how to understand the physics of a system with
$N_c \to \infty$ constituents.  A particularly simple and elegant
paradigm for dealing with this situation is the mean-field
Hartree-Fock baryon picture due to Witten\cite{Witt}, in which each
quark moves in the field caused by the other $N_c-1$ quarks acting
collectively.  To lowest order in $1/N_c$, this field is static ---
after all, a system of $N_c-1$ quarks possesses a great deal of
inertia compared to a single quark.  Then each quark in the ground
state satisfies the same field-theoretical wave equation (which, for
very heavy quarks, reduces to the Schr\"{o}dinger equation); although
the exact form of this wave equation depends upon the details of QCD
and is therefore unknown, much can be said from its existence and its
large $N_c$ scaling properties.

\begin{figure}
  \begin{centering}
  \def\epsfsize#1#2{2.7#2}
  \hfil \epsfbox{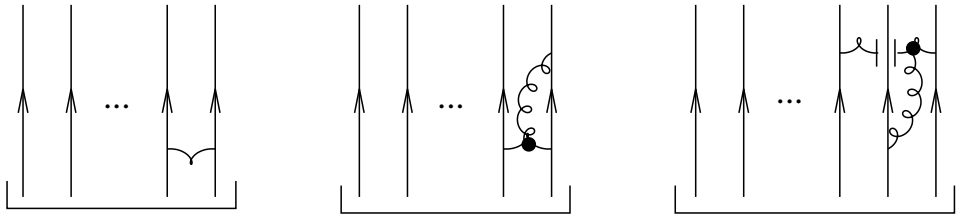}\hfil\hfil
  \vskip 0.25in

\caption{Three sample leading-order connected baryon diagrams
contributing to interaction energy.  Each square bracket indicates the
presence of $N_c$ quark lines.  Counting combinatoric factors, color
loops, and powers of $g_s \propto \sqrt{N_c}$, one finds that each
diagram scales as $N_c^1$.} \label{bary1}

  \end{centering}
\end{figure}

	For starters, since each quark in the ground state satisfies
the same wave equation, each one has the same wavefunction
$\phi(\mbox{\boldmath $r$})$, which we claim scales as $N_c^0$.  For
the scaling claim to be sensible, the potential due to the other
$N_c-1$ quarks in which each of the $N_c$ quarks moves must also scale
as $N_c^0$, or equivalently, baryon diagrams must never give
interaction energies more than $O(N_c^1)$.  To see that this is true,
one must first understand large $N_c$ counting for baryon diagrams.
The potential comes from diagrams with connected interactions such as
those depicted in Fig.~\ref{bary1}.  In addition to factors of $N_c$
from $g_s \propto 1/\sqrt{N_c}$ and color loops, one has combinatoric
factors due to the presence of $N_c$ quark lines.  For example,
diagrams involving two quarks can be chosen in ${}_{N_c}C_2 =
N_c(N_c-1)/2 = O(N_c^2)$ ways, diagrams involving three quarks can be
chosen in ${}_{N_c}C_3 = N_c(N_c-1)(N_c-2)/6 = O(N_c^3)$ ways, and so
forth.  The three diagrams in Fig.~\ref{bary1} are all seen to appear
at $O(N_c^1)$, and so are among the leading-order diagrams.  It is not
hard to prove that no diagrams are of lower order, and that internal
quark loops and nonplanar gluons (See Appendix~\ref{planar}) cost
factors of $1/N_c$, meaning that a consistent large $N_c$ counting
scheme compatible with the Hartree-Fock picture does indeed exist for
the baryons.

	Since $\phi(\mbox{\boldmath $r$})$ is the same for each quark
in the baryon, it follows that each quark has the same charge
distribution and occupies the same space; therefore, baryons remain
the same size as $N_c \to \infty$.  In contrast, large $A$ nuclei grow
in size with $A$ because of the phenomenological saturation of nuclear
forces.  Whereas large $N_c$ baryons and large $A$ nuclei both have
masses proportional to the numbers of constituents $N_c$ and $A$,
respectively, large $N_c$ baryons are like rigid containers with
radius $\propto N_c^0$ and density $\propto N_c^1$, while large $A$
nuclei are like close-packed spheres with density $\propto A^0$ and
radius $\propto A^{1/3}$.  This is why the Hartree-Fock picture works
for baryons but not nuclei.

\begin{figure}
  \begin{centering}
  \def\epsfsize#1#2{0.8#2}
  \hfil \epsfbox{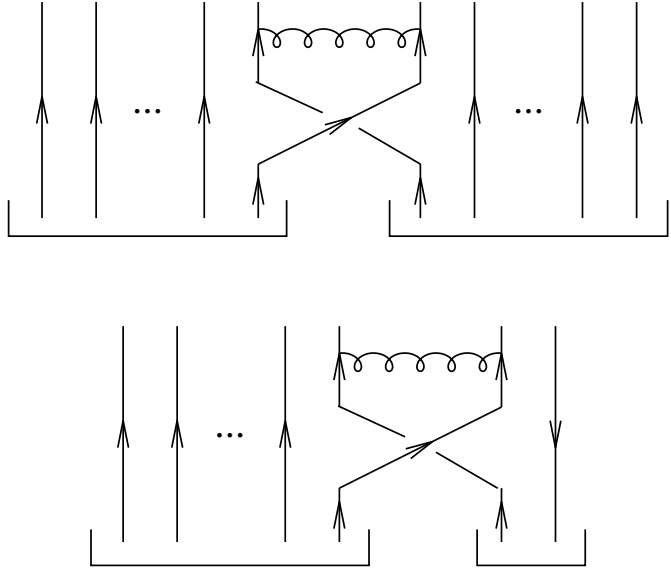}\hfil\hfil
  \vskip 0.25in

\caption{Sample $1/N_c$ leading-order diagrams for baryon-baryon and
baryon-meson scattering, respectively.  Brackets indicate quarks
collected into distinct hadrons.} \label{scat}

  \end{centering}
\end{figure}

	Within the Hartree-Fock picture it is possible to obtain large
$N_c$ behavior for common physical processes.  For example, consider
(Fig.~\ref{scat}) baryon-baryon ($BB$) and baryon-meson ($BM$)
scattering.  In the $BB$ case, one may choose from among $N_c$ quarks
in each baryon, while at least one gluon is typically required to keep
both baryons color singlets and conserve momentum, providing a factor
$(g_s/\sqrt{N_c})^2$.  The amplitude for the process thus scales as
$N_c^1$.  In the $BM$ case, the counting proceeds as before, except
that one loses one of the $N_c^1$ combinatoric factors, meaning that
the amplitude scales as $N_c^0$.  Do these factors make sense?  For
$BB$, since baryon masses are $O(N_c^1)$, an amplitude of $O(N_c^1)$
leads, via the usual textbook formulae, to a nonvanishing, finite
($O(N_c^0)$) cross-section.  For $BM$, since meson masses are
$O(N_c^0)$, the cross-section formulae lead to the scattering of the
meson --- but not the baryon --- with $O(N_c^0)$ probability; the
physical picture in this case is the meson scattering from a fixed
potential center.\footnote{To be a bit more precise about the
calculation, these results are most clear if one uses the covariant
normalization for baryons of $E/M$ particles per unit volume.
Although a mere bookkeeping device, this factor is $O(N_c^0)$ in the
large $N_c$ limit ({\it i.e.}, not infinite like $2E$) and therefore
gives the standard result for scattering from a potential center in
the $BM$ case as $N_c \to \infty$.}

	Although the Hartree-Fock picture provides a simple archetype
for the large $N_c$ baryon, its value is limited since it only
describes the leading-order $N_c$ counting for a given process.  This
was also the case for the mesons, but now with $N_c$ constituents in
the baryon, it becomes possible to perform studies of subleading
effects in $1/N_c$, which involve some subset of the quarks acting
collectively.  In the next two subsections, we shall discuss the two
known techniques for studying the subleading $1/N_c$ structure of
baryons: {\it consistency condition\/} and {\it spin-flavor algebra\/}
methods.  Both make use of an operator basis acting upon the baryon
states, and so the bulk of the $1/N_c$ suppressions is carried by the
operators.  The remainder of this lecture is dedicated to explaining
these two schemes and their phenomenological results.  The consistency
condition approach turns out to be related closely to an operator
analysis based on the Skyrme model\cite{Skyrme}, while the spin-flavor
method turns out to be closely related to operator analysis performed
in the quark model.

\subsection{Consistency Conditions} \label{cc}

\begin{figure}
  \begin{centering}
  \def\epsfsize#1#2{2.5#2}
  \hfil \epsfbox{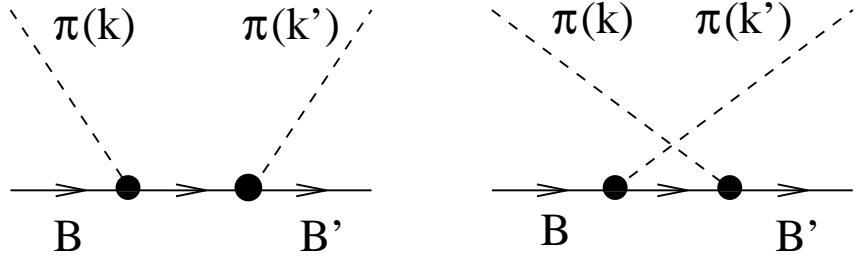}\hfil\hfil

\caption{Baryon-pion scattering diagrams used to obtain the amplitude
(\ref{conamp}), and ultimately, the consistency condition (\ref{xx}).}
\label{conscat}

  \end{centering}
\end{figure}

	The starting point of this scheme is to consider baryon-meson
scattering not in terms of quarks, but rather hadrons.  Suppose that
the meson is a $\pi$, which, as a Nambu-Goldstone boson of chiral
symmetry, couples to baryons derivatively through the isovector axial
vector current $A^{\mu a} = \bar q \gamma^\mu \gamma_5 \tau^a q$,
where $\tau$ is an isospin Pauli matrix.  The scattering process
consists of two insertions of $A^{\mu a}$ on the baryon line, which
can occur in two orderings as depicted in Fig.~\ref{conscat}.  Since
each quark in the baryon can couple to $A^{\mu a}$, one expects its
matrix elements to be $O(N_c^1)$,\footnote{The ``seagull'' diagram,
where the two pions meet at one point on the baryon, is suppressed
since it couples to the isospin quantum number of the baryon, which is
taken to be $O(N_c^0)$ for nucleons.} and the full vertex is written
as
\begin{equation}
\left< B^\prime \left| \bar q \gamma^i \gamma_5 \tau^a q \right| B
\right> \partial_i \pi^a / f_\pi \equiv \tilde{g} N_c X^{ia}_{B^\prime
B} \partial_i \pi^a / f_\pi ,
\end{equation}
where the coupling constant $\tilde{g}$ and the operator $X^{ia}$,
each of $O(N_c^0)$, are defined by this equation.  Note that only the
space components of $A^{\mu a}$ appear in the leading term of the
nonrelativistic expansion of the quark bilinear, and only these are
needed here, since at leading order in $1/N_c$ the baryon may be taken
at rest throughout the process.  This implies also that the $\pi$
scatters elastically from the baryon, and so in the first diagram in
Fig.~\ref{conscat} the baryon is off-shell by an amount $E_\pi$,
whereas in the second the amount is ($-E_\pi$).  Therefore, since the
derivatives produce factors of pion momenta $k$, the amplitude for the
two diagrams together is
\begin{equation} \label{conamp}
{\cal M} = \frac{i N_c^2 \tilde{g}^2 k^i k^{\prime \, j}}{f_\pi^2
E_\pi} \left[ X^{jb} , X^{ia} \right]_{B^\prime B} .
\end{equation}
Using that $f_\pi \propto \sqrt{N_c}$, ${\cal M}$ naively appears to
scale as $N_c^1$, rather than $N_c^0$ as concluded previously.  One
evades this conundrum by noting that many baryon states may appear on
the intermediate line, {\it i.e.}, are inserted between the two $X$'s
in the commutator; then cancellations can render the matrix elements
of the commutator $O(1/N_c)$ even though matrix elements of each $X$
individually are $O(N_c^0)$, as pointed out by Gervais and Sakita in
1984\cite{GS}.  Such cancellations are called {\it consistency
conditions}.

	The mathematics behind this approach was developed in the
1960s as one of many methods to study strong interactions, and is
generically called the method of {\it induced representations of a
contracted spin-flavor algebra}.  In order to explain this
intimidating term, we first consider what is meant by a {\it
contracted algebra}.  In essence, it means that one or more generators
of an algebra is scaled so as to change the commutation relations.
For example, perform a $1/N_c$ expansion of the operator $X$:
\begin{equation} \label{xexp}
X^{ia} = X_0^{ia} + \frac{1}{N_c} X_1^{ia} + \cdots ,
\end{equation}
in terms of which we have seen that
\begin{equation} \label{xx}
\left[ X_0^{ia} , X_0^{jb} \right] = 0.
\end{equation}
By virtue of its defining indices $i$ and $a$, $X^{ia}$ is a spin-1,
isospin-1 tensor, and therefore by definition satisfies
\begin{equation} \label{xmix}
\left[ J^i , X_0^{jb} \right] = i \epsilon^{ijk} X_0^{kb} ,
\hspace{2em}
\left[ I^a , X_0^{jb} \right] = i \epsilon^{abc} X_0^{jc} .
\end{equation}
Also, as usual,
\begin{equation} \label{su2xsu2}
\left[ J^i, J^j \right] = i \epsilon^{ijk} J^k, \hspace{2em}
\left[ I^a, I^b \right] = i \epsilon^{abc} I^c, \hspace{2em}
\left[ I^a, J^i \right] = 0 ,
\end{equation}
which are the defining relations of spin-flavor SU(2) $\times$ SU(2).
How do the commutation relations with $X_0$ alter the algebra?  To see
what happens, compare the algebra defined by
(\ref{xx})--(\ref{su2xsu2}) with that of spin-flavor SU(4).  In
addition to containing SU(2) $\times$ SU(2), SU(4) also has
commutation relations with the combined spin-flavor generator $G^{ia}
\equiv J^i \otimes I^a$,
\begin{eqnarray} \label{su4}
&& \left[ J^i , G^{ja} \right] = i \epsilon^{ijk} G^{ka} ,
\hspace{2em} \left[ I^a , G^{jb} \right] = i \epsilon^{abc} G^{ic} ,
\nonumber \\ && \left[ G^{ia}, G^{jb} \right] = \frac{i}{4} \left(
\epsilon^{ijk} \delta^{ab} J^k + \epsilon^{abc} \delta^{ij} I^c
\right) .
\end{eqnarray}
Note that the two algebras become the same if one writes
\begin{equation}
X_0^{ia} = \lim_{N_c \to \infty} G^{ia} / N_c ,
\end{equation}
and works to leading order in $1/N_c$, so that the right hand side of
the commutator $[G,G]$ vanishes.  This is the contraction of the
spin-flavor algebra.  One immediately concludes that, due to the
contraction, the smaller spin-flavor symmetry SU(2) $\times$ SU(2) is
promoted to the larger symmetry SU(4), at leading order in $1/N_c$.
For three light flavors $u$, $d$, $s$, for which the flavor symmetry
is SU(3), one finds SU(2) $\times$ SU(3) $\to$ SU(6) at leading order.
This enlargement of the symmetry due to the contraction allows one to
find a number of new relations among physical quantities.  In fact,
the SU(6) spin-flavor symmetry\cite{GR} has been known to work well
for baryons --- but rather poorly for mesons --- since the 1960s; in
large $N_c$, one sees at last a field-theoretic reason how such a
symmetry might arise.

	Next, a representation of an algebra is {\it induced\/} when
one attempts to specify the quantum numbers of a given state as
classified according to a generator of the algebra, and discovers that
other quantum numbers must be specified to label the state completely.
This happens, for example, when one classifies the irreducible
representations of the Lorentz group; once a given inertial frame
({\it i.e.}, set of boost coordinates) is chosen, one finds that there
still exists a ``little group'' of transformations needed to specify
the representation, which correspond physically to rotations in that
frame: Choosing the frame induces the full representation.  In the
present case, once one specifies eigenvalues of the operator $X_0$,
one finds\cite{CS} that there remains a ``little group'' of generators
with quantum numbers $K$, $k$, like those of angular momentum, $K =
0,1,2, \ldots$, and $k = -K, -K+1, \ldots , K-1, K$.  That is,
choosing $X_0$ induces an irreducible representation specified by the
states $\left| \right. X_0, K, k \left. \right>$.

	Of course, $X_0$ is not an intuitively familiar physical
operator, so one then projects the states $\left| \right. X_0, K, k
\left. \right>$ onto the space of states $\left| \right. I, I_3, J,
J_3, K \left. \right>$.  Certainly isospin and spin are familiar, but
what of the induced quantum number $K$?  As shown by Dashen, Jenkins,
and Manohar\cite{DJM1}, $K$ turns out to have the same properties as
the number of strange quarks in the baryon, $K = N_s/2$.  The relation
mentioned earlier between the consistency condition approach and the
Skyrme model, namely, that the group theory associated with the Skyrme
model is precisely that obtained from the consistency condition
method, is also discussed in that work.

	The consistency condition (\ref{xx}) and others like it
derived from multiple pion or kaon scattering with nucleons provide
fertile ground for phenomenological analysis.  One simply plugs an
operator possibly contributing to a given physical process into the
consistency relations, sees if it is satisfied, and if not, discards
the operator.  This process builds a $1/N_c$ expansion for a given
operator, and hence for physical quantities.  For example, one can
derive the pion-baryon coupling constant relation
\begin{equation}
g_{\pi N N}/g_{\pi N \Delta} = 3/2 + O(1/N_c^2) ,
\end{equation}
or show that $X^{ia}_1$ in (\ref{xexp}) is proportional to $X^{ia}_0$,
meaning that effectively
\begin{equation}
X^{ia} = X^{ia}_0 + O(1/N_c^2) ,
\end{equation}
so that nonstrange axial currents have no $O(1/N_c)$ corrections, or
show that the masses of nonstrange baryons of spin $J$ in the same
spin-flavor multiplet, such as the nucleons and $\Delta$'s, are first
split at relative order $J^2/N_c^2$\cite{Jenk2}, {\it i.e.},
\begin{equation}
M = N_c \left[ c_0 + c_1 \frac{J^2}{N_c^2} + O \left( \frac{1}{N_c^4}
\right) \right] ,
\end{equation}
where $c_0$ and $c_1$ are unknown coefficients of $O(N_c^0)$.  These
relations arise simply because the operators that would violate them
are disallowed by the consistency conditions.  Phenomenology seems to
agree well with this analysis; in the examples given, $g_{\pi N
\Delta}/g_{\pi N N} = 1.48$ experimentally, while the relative mass
difference between the nucleon and $\Delta$ is about $(1232-940) {\rm
MeV}/\frac 1 2 (1232 + 940) {\rm MeV} = 0.27$, which is very close to
$[J_\Delta (J_\Delta + 1) - J_N (J_N + 1) ]/N_c^2 = 0.33$ for $N_c=3$.

	A large body of additional work has been performed using
the consistency condition approach.  To sample but a few, baryon
magnetic moments\cite{JM}, orbitally excited\cite{Goity,PY}, and
hybrid\cite{CPY} have been studied in this scheme.

\subsection{Spin-flavor Algebra}

        In the consistency condition approach just discussed, we saw
that a full combined spin {\em and\/} flavor symmetry arises from the
separate spin and flavor symmetries to leading order in $1/N_c$, due
to the group contraction.  Why not, then, describe all operators
acting upon the baryons in terms of operators with definite SU($2F$)
properties, where $F$ is the number of light quark flavors, acting
upon the sum of the quarks?  That is, the operator basis in SU(6), for
example, is defined as
\begin{eqnarray} \label{su6}
J^i & \equiv & q^\dagger_\alpha \left( \frac{\sigma^i}{2} \otimes
\openone \right) q^\alpha , \nonumber \\
T^a & \equiv & q^\dagger_\alpha \left( \openone \otimes
\frac{\lambda^a}{2} \right) q^\alpha , \nonumber \\
G^{ia} & \equiv & q^\dagger_\alpha \left( \frac{\sigma^i}{2} \otimes
\frac{\lambda^a}{2} \right) q^\alpha ,
\end{eqnarray}
where $\sigma^i$ are the usual Pauli spin matrices, $\lambda^a$
denotes Gell-Mann flavor matrices, and the index $\alpha$ sums over
all $N_c$ quark lines in the baryons.  One can readily show that $J$,
$T$, and $G$ satisfy an SU(6) algebra analogous to the SU(4) algebra
Eq.~(\ref{su4}).\footnote{To be precise, one replaces $I^a \to T^a$,
$\epsilon^{abc} \to f^{abc}$, and the $[G,G]$ commutator becomes
\begin{equation} \label{gg}
[G^{ia}, G^{jb} ] = \frac{i}{4} \delta^{ij} f^{abc} T^c + \frac{i}{6}
\delta^{ab} \epsilon^{ijk} J^k + \frac{i}{2} \epsilon^{ijk} d^{abc}
G^{kc} .
\end{equation}}

	An important point to realize is that, although baryons are
described in this approach in terms of $N_c$ ``quarks,'' a large $N_c$
constituent quark model\cite{KP} is not assumed.  Indeed, recalling
the lesson from Sec.~\ref{mes} that valence and hybrid mesons mix
freely, one might well expect the same for hybrid baryons\cite{CPY}.
However, the only assumption really used to perform this operator
analysis is that baryons for large $N_c$ continue to have the same
total quantum numbers as those obtained from an $N_c$ quark system.
Then, a ``quark'' in the sense of Eq.~(\ref{su6}) is not the same as
the dynamical entity of the QCD Lagrangian, but rather a
group-theoretical object representing one part in $N_c$ of the
physical baryon.  The two definitions coincide when the quarks are
taken very heavy, so that their interactions with the internal gluons
become less important in determining the baryon structure.

	It was shown by Manohar\cite{Man2} that results of the Skyrme
model obtained from its operator structure (and hence obtained from
the consistency condition approach) and those obtained from the
operator structure of the quark model (and hence obtained from the
spin-flavor algebra approach) coincide as $N_c \to \infty$.
Therefore, either approach is equally good from the mathematical point
of view for describing large $N_c$ baryons.  Moreover, the total
result for any given quantity must be the same regardless of which
approach one uses, which is simply the statement that the physical
result must be independent of mathematics.  One sees that the
consistency condition and spin-flavor algebra methods differ for a
given physical quantity only by a reorganization of $1/N_c$
corrections.

	The initial work on the baryon $1/N_c$ expansion in the
spin-flavor algebra approach was performed by Carone, Georgi, and
Osofsky\cite{CGO} and Luty and March-Russell\cite{LMR}.  The basic
results of these analyses is that many of the old SU(6) static
relations between baryonic matrix elements are recovered by means of
an operator expansion in $1/N_c$.  That is, any given physical
operator ${\cal O}^{(m)}$ that scales as $N_c^m$ may be written in the
form
\begin{equation} \label{expan}
{\cal O}^{(m)} = N_c^m \sum_{n,p,q} c_n \biggl( \frac{J^i}{N_c}
\biggr)^p \biggl( \frac{T^a}{N_c} \biggr)^q \biggl( \frac{G^{jb}}{N_c}
\biggr)^{n-p-q} ,
\end{equation}
where it is understood that spin and flavor indices are to be
contracted in such a way as to agree with the transformation
properties of ${\cal O}^{(m)}$.

	The factor of $1/N_c$ accompanying each $J$, $T$, and $G$
reflects the fact that the matrix elements of these operators can
often add {\em coherently\/} over the $N_c$ quarks and thus can be as
large as $O(N_c^1)$.  In fact, this was the central problem with the
spin-flavor algebra method: One hopes to describe a given quantity
with some small number of operators, but if a number of the operators
in (\ref{expan}) have matrix elements of $O(N_c^1)$, where can it
justifiably be truncated?  Before addressing this problem directly, it
is useful to consider the nature of the baryon multiplets in this
approach.

	Since we are making use of representations of the symmetry
group SU($2F$) for operators acting upon the baryons, the most
convenient description for the baryons is in terms of multiplets of
SU($2F$).  Recall that this is the symmetry of independent rotations
in all spin SU(2) and flavor SU($F$) coordinates, under which the
ground state baryons appear to fill a completely symmetric
representation.  It is very convenient to describe the representations
in terms of Young tableaux, in which each quark, which appears in a
fundamental representation of SU($2F$), is represented by a single
box.  Boxes arranged in a row indicate symmetrization of the
corresponding SU($2F$) indices, while boxes arranged in a column
indicate antisymmetrization.  The tableau for the completely symmetric
SU($2F$) multiplet of $N_c$ quarks is a row of $N_c$ boxes, as in the
first line of Fig.~\ref{young}; using standard group theory
techniques, this representation can be decomposed into a tower of
component ${\rm SU}(F)_{\rm flavor} \times {\rm SU(2)}_{\rm spin}$
representations, as indicated in the succeeding lines of
Fig.~\ref{young}.  Each Young tableau in this tower has $N_c$ boxes.
If you are familiar with the old SU(6) theory of baryons, then you
will recognize the special case of $N_c=3$, $F=3$ as the decomposition
\begin{equation}
{\bf 56} = \left( {\bf 8}, J= \frac 1 2 \right) \oplus \left( {\bf
10}, J= \frac 3 2 \right) .
\end{equation}
The significance of this result is twofold: First, the number of
SU($F$) $\times$ SU(2) representations increases as $N_c$ is increased
from its physical value of 3, and second, the individual SU($F$)
multiplets themselves become much larger as $N_c$ is increased beyond
three.\footnote{There is one exception, namely, $F=2$.  In that case,
all of the two-box columns become isosinglets and may be ignored, and
then the isomultiplet representations are independent of $N_c$ ---
indeed, they satisfy $I=J$.}  This point is illustrated in
Figs.~\ref{spin12} and \ref{spin32} for the case of three light
flavors; note how the familiar SU(3) octet (which contains the
nucleons) and the decuplet (which contains the $\Delta$ resonances)
are obtained from shrinking the representation down to $N_c=3$.

	Let us now return to the conundrum of where to truncate the
expansion (\ref{expan}).  We see that working with $N_c$ large
generates an enormous number of baryon states, but we are only
interested in the phenomenology of those with which we are familiar,
{\it i.e.}, that already exist for $N_c=3$, which we denote {\it
physical\/} baryons.  Where do they appear within the large multiplets
of Figs.~\ref{spin12}, \ref{spin32}?  The mathematics alone allows for
a great deal of freedom for placing the physical states, but the
usefulness of (\ref{expan}) as a perturbative expansion in $1/N_c$
imposes constraints on this choice.  For example, since the expansion
breaks down if $J=O(N_c^1)$ or $I=O(N_c^1)$, we take physical baryons
to have $J=O(N_c^0)$ and $I=O(N_c^0)$.  As for the strangeness
content, consider for example the top-right entry of the multiplet
Fig.~\ref{spin12}; its spin-flavor structure is $uud$ in a total
spin-1/2, isospin-1/2 state (just as for an $N_c=3$ proton) plus
$(N_c-3)/2$ $ud$ pairs, each combined into spin-0, isospin-0 quantum
numbers.  It is tempting to identify this state as the $N_c>3$
analogue of the proton, and one may do this by fiat; indeed, in light
of our statements above, the physical proton should have $J=I=1/2$
even for $N_c>3$.  Apart from the top-right state, the physical proton
might occupy any of a number of sites in Fig.~\ref{spin12} that lie
directly below this one.  However, each of these states contains a
number of $s$ quarks, which are heavier than $u$ and $d$ quarks, and
therefore these baryons should be heavier than those on the top row.
If the physical proton contains some valence $s$ quarks, it decays
weakly to some other lighter baryon, a process that opens up a
phenomenological Pandora's box.  One does not encounter any of these
or similar problems if one identifies the physical baryons to lie at
the tops of Figs.~\ref{spin12} and \ref{spin32}.

	Even taking these assignments into account, some of the
operators, such as $T^{4,5,6,7}$ and $G^{ia}$, can still give matrix
elements of $O(N_c)$, even when evaluated only on the physical
baryons.  All is not lost, however, since we have not to this point
made full use of the mathematics.  Since the operators can be thought
of as a set of vectors spanning the space of physical observables, it
must be that there are only as many independent operators as
independent observables.  One can obtain a number of {\it operator
reduction rules}\cite{DJM2} based upon both the structure of the
spin-flavor algebra SU($2F$) and properties particular to the
completely symmetric ground-state representation.

	To see how this works, collect the SU($2F$) generators into a
uniformly normalized set:
\begin{equation} \label{collect}
\Lambda^A \equiv \left\{ \left( \sigma^i/2 \otimes \openone \right)
/\sqrt{F}, \, \left( \openone \otimes \lambda^a/2 \right) /\sqrt{2},
\, \sqrt{2} \left( \sigma^i/2 \otimes \lambda^a/2 \right) \right\} \to
{\rm Tr \,} \Lambda^A \Lambda^B = \frac 1 2 \delta^{AB} .
\end{equation}
Then two simple operator reduction relations are
\begin{eqnarray}
q^\dagger_\alpha ( \openone \otimes \openone ) q^\alpha & = & N_c
\openone , \label{orr1} \\ \left( q^\dagger_\alpha \, \Lambda^A
q^\alpha \right) \left( q^\dagger_\beta \, \Lambda^A q^\beta \right) &
= & \frac{N_c}{2} (N_c + 2F) \left( 1 - \frac{1}{2F} \right)
\openone . \label{orr2}
\end{eqnarray}
Equation (\ref{orr1}) is actually nothing more than the total quark
number operator in the baryon, which of course produces a factor
$N_c$.  The left-hand side of Eq.~(\ref{orr2}) by definition is the
quadratic Casimir of SU($2F$) in the same way that $J^2$ is the
Casimir of spin SU(2), and therefore is also proportional to the
identity operator.  Note in both cases how each reduction by one
operator leads to an additional power of $N_c$; it is this feature
that maintains the consistency of large $N_c$ counting within the
operator reduction rules.  Once the complete set of such rules is
known, one may produce a minimal set of independent operators to any
given order in the $1/N_c$ expansion to describe a given physical
quantity.

	Analysis of the mass spectrum~\cite{JL} provides a good
illustration of these ideas.  Let us consider here for simplicity only
the eight isospin-averaged masses $N$, $\Sigma$, $\Xi$, $\Delta$,
$\Sigma^*$, $\Xi^*$, $\Omega$ of the ground-state spin-flavor
multiplet, where the baryon label stands for its mass.  For $N_c > 3$
the relevant baryon multiplets contain many additional states, as a
glance at Figs.~\ref{spin12}, \ref{spin32} verifies, but for sake of
phenomenology only these 8 physical baryons need be considered.  One
must therefore generate the leading 8 independent operators in the
$1/N_c$ expansion, taking into account all operator reduction rules,
in order to obtain a set of mass operators that forms a complete basis
(in the sense of a vector space) for spanning all possible values of
masses.  The operators with attendant $1/N_c$ factors are
\begin{equation} \label{ops}
N_c \openone, \ T^8, \ \frac{1}{N_c} J^2, \ \frac{1}{N_c} J^i G^{i8},
\ \frac{1}{N_c} \left( T^8 \right)^2, \ \frac{1}{N_c^2} J^2 T^8, \
\frac{1}{N_c^2} J^i G^{i8} T^8, \ \frac{1}{N_c^2} \left( T^8 \right)^3
.
\end{equation}
The mass Hamiltonian operator is simply a linear combination of these
operators, each with an arbitrary coefficient.  One may now compute
the matrix element of each operator for each baryon, obtaining in this
way an expression for each coefficient in terms of baryon masses.  For
example, the coefficient of $(T^8)^2$ is fixed by the combination
\begin{equation} \label{mcomb}
35 ( 2N - \Sigma - 3\Lambda + 2\Xi ) - 4 (4\Delta - 5\Sigma^* - 2\Xi^*
+ 3\Omega ) .
\end{equation}
Our mass Hamiltonian has not yet taken into account all the known
physics of the ground-state multiplet; in particular, the masses of
the baryons are known to break along the hypercharge (strangeness)
axis, which corresponds to the explicit flavor = 8 indices appearing
in (\ref{ops}).  This explicit SU(3) breaking is known to be
relatively small, at the order of $\epsilon \approx$ 0.25--0.30, since
the relative mass difference between, say, $N (940)$ and $\Sigma
(1190)$ is roughly this size.  Every time a flavor = 8 index appears,
let us include a factor of $\epsilon$ in the mass Hamiltonian.

\begin{figure}
  \begin{centering}
  \def\epsfsize#1#2{0.6#2}
  \hfil \epsfbox{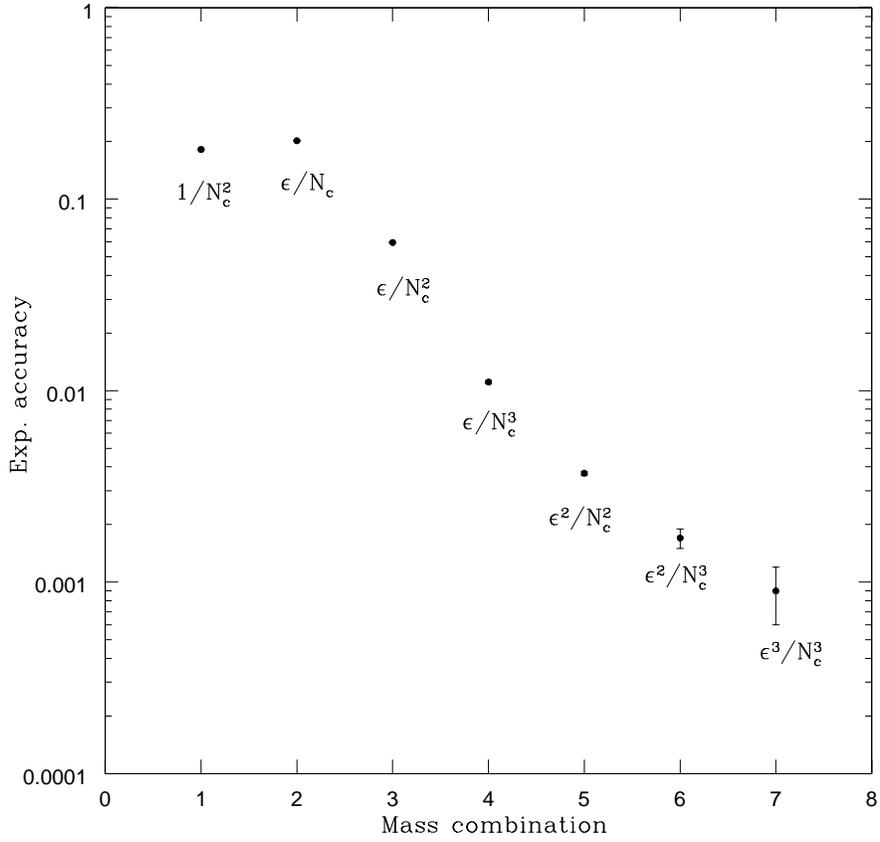}\hfil\hfil

\caption{Experimental values of isospin-averaged mass combinations
relative to the common baryon mass, as described in the text (a ratio
called experimental accuracy here).  Each one corresponds to a
particular operator in the $1/N_c$ expansion, whose $1/N_c$ and
SU(3)-breaking suppressions $\epsilon$ label each point.}
\label{mass}

  \end{centering}
\end{figure}

	One would now like to compare the matrix element $\epsilon^2
\left< (T^8)^2 \right>/N_c$ to the mass combination (\ref{mcomb}), but
both of these expressions still possess scale independence.  That is,
both (\ref{mcomb}) and the coefficients of the mass Hamiltonian may be
multiplied by arbitrary fixed numbers.  One overcomes this ambiguity
by comparing to the $O(N_c^1)$ common mass of the baryons.  On the
operator side in our example, this means that the relevant quantity is
$\epsilon^2/N_c^2$, which for $\epsilon = 0.25$ and $N_c = 3$ is about
0.69\%.  On the mass side, one organizes a given linear combination of
masses into the form LHS = RHS, where all numerical coefficients on
either side are positive, and since the combination vanishes in the
spin-flavor symmetry limit, the total of the LHS and RHS coefficients
are equal.  One then forms the combination (LHS + RHS)/2.  In our
example, this is
\begin{equation}
\frac{1}{2} \left[ 35 (2N + \Sigma + 3\Lambda + 2\Xi) + 4 (4\Delta +
5\Sigma^* + 2\Xi^* + 3\Omega ) \right] .
\end{equation}
The desired scale-independent combination is then $|{\rm LHS}-{\rm
RHS}|/\frac{1}{2}$(LHS+RHS).  Here, this is experimentally $0.37 \pm
0.01 \%$; thus the unknown coefficient of the operator $\epsilon^2
(T^8)^2 / N_c$ has the very natural size 0.53.  Had we ignored the
factors of $N_c$, the coefficient would have been 0.06, its small size
suggesting that some important physical suppression had been ignored.
Carrying out this analysis for all 8 masses gives the plot in
Fig.~\ref{mass}; here one sees graphically that not only the factors
$\epsilon$ but also the factors of $N_c = 3$ are needed to understand
satisfactorily the mass spectrum.

	The truly interesting feature of this analysis is that large
$N_c$ is used to obtain the relevant operator expansion represented by
(\ref{ops}), but the phenomenological analysis then employs $N_c=3$.
One concludes that, at least for this case, the methods of large $N_c$
seem to survive extrapolation all the way to the rather small observed
value of $N_c=3$.  Similar analysis in the quark operator
representation has been performed for the magnetic
moments\cite{LMRW,DDJM} and axial-vector couplings\cite{DDJM,FJM},
where evidence for the predictions of the $1/N_c$ expansion is found.

	However, this is not to say that the $1/N_c$ expansion is
perfect!  An interesting example of a case where the expansion for
$N_c=3$ does not explain everything is that of nonstrange $\ell = 1$
excited baryons\cite{CCGL}, such as the $N(1535)$.  An analysis
similar to that described above (somewhat more complicated due to the
additional presence of the orbital angular momentum operator $\ell^i$)
shows that a number of the mass Hamiltonian coefficients have central
fit values that are quite small, albeit with substantial
uncertainties.  It is not yet clear whether this is due entirely to
large experimental uncertainties in the mass determination of these
resonances,\footnote{However, data from CLAS at Jefferson Lab should
improve upon the experimental values.} or additional dynamics beyond
the $1/N_c$ expansion, or both.  On the other hand, if any of the
coefficients had turned out too large, it would have been an
unmitigated failure of large $N_c$, since in that case the claimed
$1/N_c$ suppressions simply would not be supported by experiment.

\subsection{Nuclear Physics}

	Large $N_c$ nuclear theory is a subfield still in its infancy,
but the few results obtained thus far are encouraging.  Interestingly,
all of the current results tend to follow from the observation that
$G^{ia}$ has $O(N_c^1)$ matrix elements for the physical baryons,
while those of $I^a$ and $J^i$ are $O(N_c^0)$.  In Ref.~\cite{KS}, the
dominance of $G^{ia}$ is used to show that the $1/N_c$ leading-order
central nucleon-nucleon (NN) interaction is SU(4) symmetric.  To see
this, consider the collected SU(4) generators $\Lambda^A$, as defined
in Eq.~(\ref{collect}).  The leading SU(4) symmetric operator for the
system consisting of nucleons 1 and 2 is
\begin{equation}
\left[ \Lambda^A \right]_1 \left[ \Lambda^A \right]_2 ,
\end{equation}
which contains
\begin{equation}
\left[ G^{ia} \right]_1 \left[ G^{ia} \right]_2 + O(1/N_c^2),
\end{equation}
where the $O(1/N_c^2)$ corrections come from $I_1 \cdot I_2$ and $J_1
\cdot J_2$ terms.  Since $G^{ia}$ is a leading-order operator, one
concludes that the leading-order interaction is SU(4) symmetric.  This
reasoning is very similar to that of Sec.~\ref{cc}, where group
contraction promotes separate SU(2)$_{\rm spin}$ and SU(2)$_{\rm
isospin}$ symmetries to SU(4), since it depends upon terms in $[G,G]$
(Eq.~(\ref{gg})) being subleading in $1/N_c$.

	One very interesting consequence of this result is an
explanation for the observed Wigner supermultiplet symmetry.  Many
nuclear interactions seem to obey a symmetry among the states in the
multiplet
\begin{equation}
\left( p \uparrow, \, p \downarrow, \, n \uparrow, \, n \downarrow
\right),
\end{equation}
and large $N_c$ provides an explanation\cite{KS}: Given the
NN Lagrangian only up to dimension-6 operators, one has
\begin{equation}
{\cal L}_6 = -\frac{1}{2} C_S \left( N^\dagger N \right)^2 -
\frac{1}{2} C_T \left( N^\dagger \sigma^i N \right)^2 ,
\end{equation}
where the Wigner symmetry is preserved by the first term and broken by
the second.  However, the second breaks SU(4) and therefore is
subleading in $1/N_c$.

	The leading $1/N_c$ behavior of the non-central terms is
derived in Ref.~\cite{KM}, where the nonrelativistic, elastic NN
potential is expressed through the decomposition
\begin{eqnarray}
V_{\rm NN} & = & \left[ V_0^0 + V_\sigma^0 \, \mbox{\boldmath
$\sigma_1 \cdot \sigma_2$} + V_{LS}^0 \, \mbox{\boldmath $L \cdot S$}
+ V_T^0 \, S_{12} + V_Q^0 \, Q_{12} \right] \nonumber \\ & + & \left[
V_0^1 + V_\sigma^1 \, \mbox{\boldmath $\sigma_1 \cdot \sigma_2$} +
V_{LS}^1 \, \mbox{ \boldmath $L \cdot S$} + V_T^1 \, S_{12} + V_Q^1 \,
Q_{12} \right] \mbox{\boldmath $\tau_1 \cdot \tau_2$},
\end{eqnarray}
where
\begin{eqnarray}
S_{12} & \equiv & 3 \, \mbox{\boldmath $\sigma_1 \cdot \hat r \, \sigma_2
\cdot \hat r$ $-$ $\sigma_1 \cdot \sigma_2$} , \nonumber \\ Q_{12} &
\equiv & \frac{1}{2} \left\{ \mbox{\boldmath $\sigma_1 \cdot L$}, \,
\mbox{\boldmath $\sigma_2 \cdot L$} \right\} .
\end{eqnarray}
Again using the dominance of $G^{ia}$, one can show that the potential
terms $V_0^0$, $V_\sigma^1$, and $V_T^1$ appear at $O(N_c^1)$, while
every other term is relatively suppressed by at least two powers of
$1/N_c$.  The immediate consequence for the NN interaction is that
these terms should dominate the phase shifts, and indeed this is the
case.  As has been long known, the conspicuous features of the NN
interaction are a spin-singlet central potential, for which $V^0_0$ is
prominent, and an isotriplet Lorentz tensor force, for which
$V^1_{\sigma}$ and $V^1_T$ are prominent.  In phenomenologically
successful one-meson exchange potential models these features are
represented by including large coupling constants for $\pi$, $\sigma$,
and $\omega$ mesons for the central force, and $\pi$ and $\rho$ for
the tensor force.  However, the salient point is that the relative
importance of these terms in the potential derives directly from large
$N_c$ counting, regardless of the physical origin of these couplings.
Any phenomenologically successful model must obey the same pattern.

\subsection{Further Advances}

	Phenomenological studies of the $1/N_c$ expansion comprise an
increasingly large volume of literature.  While these notes are
designed as lectures rather than a complete review, I would like at
least to mention a number of additional interesting problems that have
been tackled to date.

\begin{itemize}

	\item That the $1/N_c$ expansion and chiral perturbation
theory of baryons may be combined was demonstrated by
Jenkins\cite{Jenk3}.  The feature discussed earlier that mesons appear
in nonets rather than octets of SU(3) flavor ({\it i.e.}, that the
$\eta^\prime$ meson should be included in the chiral Lagrangian) is
supported by this construction.  One finds in computing loops with
baryon lines in this Lagrangian (which requires including towers of
intermediate states: See, {\it e.g.}, Fig.~\ref{conscat}) that the
tree-level counting of powers of $1/N_c$ is maintained; that is, loops
produce complicated functions of the chiral symmetry breaking
parameters only (such as $m_\pi^2 \ln m_\pi^2$), not $N_c$.

	\item The famous Heavy Quark Effective Theory\cite{Neu} (HQET)
relates physical amplitudes of hadrons with heavy quarks ($b$ or $c$)
of different spins or flavors based on the idea that a sufficiently
heavy quark is like a static color source, the cloud of light quarks
and gluons surrounding it barely noticing its particular spin or
flavor quantum numbers.  The relation of different amplitudes, in
particular form factors, means that an approximate symmetry is at work
in HQET; however, since large $N_c$ induces an approximate symmetry as
well (spin-flavor SU($2F$)), one would expect that large $N_c$ and
HQET together produce additional constraints.  This is indeed the
case, as shown by Chow\cite{Chow}: There are relations between the
form factors for the semileptonic decays $\Lambda_b \to \Lambda_c \ell
\bar \nu$, $\Sigma_b \to \Sigma_c \ell \bar \nu$, and $\Sigma^*_b \to
\Sigma^*_c \ell \bar \nu$.

	\item Large $N_c$ and HQET can be combined into a larger and
highly predictive effective theory, as shown by Jenkins\cite{Jenk4}.
The large $N_c$ spin-flavor symmetry in this case is SU($2F_{\rm
heavy}$) $\times$ SU($2F_{\rm light}$).\footnote{For comparison, the
$F_{\rm heavy} = 1$ case was explored in \cite{Lebed}.}  One obtains
from this study a hierarchy of charmed baryon masses, in analogy to
that of \cite{JL}, whose masses agree with experimental values.  One
can then use the charmed masses to predict the yet unmeasured beauty
baryon masses to high accuracy.

	\item The decays of $\ell=1$ baryons to ground-state baryons
through pions\cite{CGKM} and through photons\cite{CC} have been
investigated in the $1/N_c$ expansion.  These decays will be measured
in great numbers at Jefferson Lab.

\end{itemize}

	This sampling of different directions of research should
demonstrate that the field of large $N_c$ phenomenology is quite vital
and holds the promise of producing many further interesting, relevant
results.  I invite you to add to the sum of this knowledge.

\section{Conclusions}

	Large $N_c$ QCD is an elegant theoretical construct that
organizes all QCD Feynman diagrams into a countable set of classes
based on their topological properties.  The $1/N_c$ expansion thus
obtained is the only known way to turn QCD into a truly perturbative
theory at all energies.  Predictions of large $N_c$ are obtained by
counting the numbers of powers of $N_c$ associated with each diagram;
since this hierarchy persists when infinite numbers of Feynman
diagrams are summed to give diagrams for hadronic processes, the
approach graduates from one of mathematical curiosity to real
phenomenological relevance.

	For mesons, many phenomenologically observed but poorly
understood properties simply follow from the leading-order term of the
$1/N_c$ expansion, while the operator expansion for baryons provides a
vehicle for systematic quantitative analysis of masses, couplings, and
other physical observables.  This is an active area of current
research at the time of this writing.

	All of the results presented here, with the exception of the
't~Hooft model in Sec.~\ref{mes}, rely solely on the large $N_c$
counting rules.  Essentially, the counting rules provide nothing more
than an organizing principle for contributions to physical quantities,
while the complicated details of the dynamics reside in the
coefficients of these operators.

	Considering the numerous successes of large $N_c$, it seems
altogether possible that the eventual solution of QCD will depend
intimately on this remarkable limit.  This is an opinion shared by
many of those who perform much more formal field-theoretical work and
are endeavoring to compute the dynamical coefficient in from of the
leading term in $1/N_c$ --- that is, to sum all the planar graphs.
Can this be done for real QCD?  If so, perhaps buried in the result of
this important calculation are treasures like the origins of
confinement and chiral symmetry breaking.

\section*{Acknowledgments}

Many thanks to the organizers and students for making the summer
school a success.  I would especially like to acknowledge C. Carlson,
C. Carone, and J. Goity for comments on this manuscript.  This work
was supported by the U.S. Department of Energy under contract No.\
DE-AC05-84ER40150.

\appendix
\section{``Planarity'' in Baryon Diagrams} \label{planar}

\begin{figure}
  \begin{centering}
  \def\epsfsize#1#2{1.3#2}
  \hfil \epsfbox{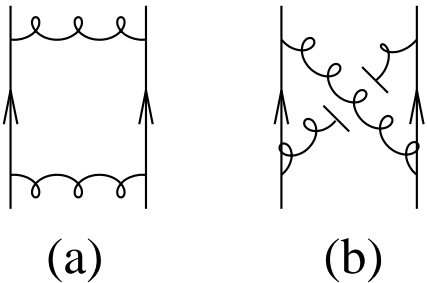}\hfil\hfil

\caption{Simple ladder ($a$) and crossed-ladder ($b$) gluon exchange
diagrams for baryons.} \label{ladd}

  \end{centering}
\end{figure}

	The question of what constitutes a ``nonplanar'' gluon for
baryons is more tricky for baryons than mesons.  Consider
Fig.~\ref{ladd}; in Fig.~\ref{ladd}$a$ one sees an ordinary ``ladder''
diagram that for the mesons would be planar, whereas the crossed
diagram Fig.~\ref{ladd}$b$ would be nonplanar.  For baryons, however,
the double-line notation shows that $a$ has no closed color loops,
whereas $b$ has one.  Therefore, since both diagrams share the
combinatoric factor ${}_{N_c}C_2$ and four factors of $g_s$, the two
diagrams scale respectively as $N_c^0$ and $N_c^1$;
Figure~\ref{ladd}$b$ is therefore a leading-order diagram.  Indeed, we
leave it as an exercise for the reader to draw the color flow for the
two diagrams, in order to be convinced that maintaining the direction
of the arrows on the color lines requires the double lines of gluons
to be twisted.  Since the two diagrams differ by the crossing of a
gluon line, one may call $a$ ``nonplanar'' and $b$ ``planar''.

	Why has this complication occurred?  The key is that all of
the quarks in the baryon have the same direction of color flow.  For
mesons, quarks and antiquarks (which have opposite directions of color
flow) always appear in pairs, and therefore adjacent color loops in
the double-line notation always have lines pointing in opposite
directions.  This is precisely what is needed to produce orientable
surfaces (see Footnote~\ref{orient}) and a convenient diagrammatic
topological classification of meson diagrams, and what is lacking in
the baryon case.  Whereas for mesons each nonplanarity cost a factor
of $1/N_c^2$, we see from our example that nonplanar $1/N_c^1$
suppressions for baryons are possible.

\def\nboxA{\vbox{\hbox{$\bsqr\bsqr\bsqr\bsqr\raise2.7pt\hbox{$\,\cdot
\cdot\cdot\cdot\cdot\,$}\bsqr\bsqr$}\nointerlineskip 
\kern-.3pt\hbox{$\bsqr$}}}

\def\nboxE{\vbox{\hbox{$\bsqr\bsqr\bsqr\raise2.7pt\hbox{$\,\cdot\cdot
\cdot\cdot\cdot\,$}\bsqr\bsqr\bsqr\bsqr$}\nointerlineskip 
\kern-.2pt\hbox{$\bsqr\bsqr\bsqr\raise2.7pt\hbox{$\,\cdot\cdot\cdot
\cdot\cdot\,$}\bsqr$}}}

\def\nboxF{\vbox{\hbox{$\bsqr\bsqr\bsqr\bsqr\raise2.7pt\hbox{$\,\cdot
\cdot\cdot\cdot\cdot\,$}\bsqr\bsqr$}\nointerlineskip 
\kern-.2pt\hbox{$\bsqr\bsqr\bsqr\bsqr\raise2.7pt\hbox{$\,\cdot\cdot
\cdot\cdot\cdot\,$}\bsqr$}}}

\begin{figure}
  \begin{raggedright}

\def\ssqr#1#2{{\vbox{\hrule height #2pt
      \hbox{\vrule width #2pt height#1pt \kern#1pt\vrule width #2pt}
      \hrule height #2pt}\kern- #2pt}}
\def\sqr{\mathchoice\ssqr8{.4}\ssqr8{.4}\ssqr{5}{.3}\ssqr{4}{.3}}

\def\bsqr{\ssqr{15}{.3}}

\def\nbox{\hbox{$\bsqr\bsqr\bsqr\bsqr\raise7pt\hbox{$\,\cdot\cdot\cdot
\cdot\cdot\,$}\bsqr\bsqr\bsqr$}}

\centerline{$$\nbox$$\mbox{\hspace{2em}}}
\bigskip\bigskip
\centerline{\raise12pt\hbox{= \hspace{1em}} $$\nboxF$$
\raise10pt\hbox{, \hspace{1em} $J= \frac 1 2$}}
\bigskip
\centerline{\raise12pt\hbox{$\oplus$ \hspace{1em}} $$\nboxE$$
\raise10pt\hbox{, \hspace{1em} $J= \frac 3 2$}}
\bigskip
\centerline{\hspace{-1em}$\cdot$}
\centerline{\hspace{-1em}$\cdot$}
\centerline{\hspace{-1em}$\cdot$}
\bigskip
\centerline{\raise5pt\hbox{$\oplus$ \hspace{1em}} $$\nbox$$
\raise5pt\hbox{, \hspace{1em} $J= \frac{N_c}{2}$}}
\bigskip

\caption{The completely symmetric SU($2F$) $N_c$-box Young tableau,
corresponding to ground-\-state baryons, decomposed into SU($F$)$_{\rm
flavor}$ $\times$ SU(2)$_{\rm spin}$ representations.}
\label{young}

  \end{raggedright}
\end{figure}

\def\onedot{\makebox(0,0){$\scriptstyle 1$}}
\def\twodot{\makebox(0,0){$\scriptstyle 2$}}
\def\threedot{\makebox(0,0){$\scriptstyle 3$}}
\def\fourdot{\makebox(0,0){$\scriptstyle 4$}}

\setlength{\unitlength}{3mm}

\begin{figure}
  \begin{centering}

\centerline{\hbox{
\begin{picture}(20.79,18)(-10.395,-8)
\multiput(-1.155,10)(2.31,0){2}{\onedot}
\multiput(-2.31,8)(4.62,0){2}{\onedot}
\multiput(-3.465,6)(6.93,0){2}{\onedot}
\multiput(-4.62,4)(9.24,0){2}{\onedot}
\multiput(-5.775,2)(11.55,0){2}{\onedot}
\multiput(-6.93,0)(13.86,0){2}{\onedot}
\multiput(-8.085,-2)(16.17,0){2}{\onedot}
\multiput(-9.24,-4)(18.48,0){2}{\onedot}
\multiput(-10.395,-6)(20.79,0){2}{\onedot}
\multiput(-9.24,-8)(2.31,0){9}{\onedot}
\multiput(0,8)(2.31,0){1}{\twodot}
\multiput(-1.155,6)(2.31,0){2}{\twodot}
\multiput(-2.31,4)(2.31,0){3}{\twodot}
\multiput(-3.465,2)(2.31,0){4}{\twodot}
\multiput(-4.62,0)(2.31,0){5}{\twodot}
\multiput(-5.775,-2)(2.31,0){6}{\twodot}
\multiput(-6.93,-4)(2.31,0){7}{\twodot}
\multiput(-8.085,-6)(2.31,0){8}{\twodot}
\end{picture}
}}

\bigskip

\caption{Weight diagram corresponding to the ground-state spin-1/2
representation of SU(3) for $N_c > 3$.  Indicated are the numbers of
states at a given site.  The longer side has $(N_c+1)/2$ sites.}
\label{spin12}

  \end{centering}
\end{figure}

\begin{figure}
  \begin{centering}
\centerline{\hbox{
\begin{picture}(20.79,18)(-8.085,-8)
\multiput(-1.155,10)(2.31,0){4}{\onedot}
\multiput(-2.31,8)(9.24,0){2}{\onedot}
\multiput(-3.465,6)(11.55,0){2}{\onedot}
\multiput(-4.62,4)(13.86,0){2}{\onedot}
\multiput(-5.775,2)(16.17,0){2}{\onedot}
\multiput(-6.93,0)(18.48,0){2}{\onedot}
\multiput(-8.085,-2)(20.79,0){2}{\onedot}
\multiput(-6.93,-4)(18.48,0){2}{\onedot}
\multiput(-5.775,-6)(16.17,0){2}{\onedot}
\multiput(-4.62,-8)(2.31,0){7}{\onedot}
\multiput(0,8)(2.31,0){3}{\twodot}
\multiput(-1.155,6)(6.93,0){2}{\twodot}
\multiput(-2.31,4)(9.24,0){2}{\twodot}
\multiput(-3.465,2)(11.55,0){2}{\twodot}
\multiput(-4.62,0)(13.86,0){2}{\twodot}
\multiput(-5.775,-2)(16.17,0){2}{\twodot}
\multiput(-4.62,-4)(13.86,0){2}{\twodot}
\multiput(-3.465,-6)(2.31,0){6}{\twodot}
\multiput(1.155,6)(2.31,0){2}{\threedot}
\multiput(0,4)(4.62,0){2}{\threedot}
\multiput(-1.155,2)(6.93,0){2}{\threedot}
\multiput(-2.31,0)(9.24,0){2}{\threedot}
\multiput(-3.465,-2)(11.55,0){2}{\threedot}
\multiput(-2.31,-4)(2.31,0){5}{\threedot}
\multiput(2.31,4)(2.31,0){1}{\fourdot}
\multiput(1.155,2)(2.31,0){2}{\fourdot}
\multiput(0,0)(2.31,0){3}{\fourdot}
\multiput(-1.155,-2)(2.31,0){4}{\fourdot}
\end{picture}
}}

\bigskip

\caption{Weight diagram corresponding to the ground-state spin-3/2
representation of SU(3) for $N_c > 3$.  Indicated are the numbers of
states at a given site.  The longer side has $(N_c-1)/2$ sites.}
\label{spin32}

  \end{centering}
\end{figure}

\end{document}